\documentclass[amsmath,amssymb,
superscriptaddress,nofootinbib]{JHEP3} % 10pt is ignored!

\usepackage{graphicx}
\usepackage{epsfig,multicol,bbm}
\usepackage{amsbsy}
\usepackage{amsmath, amsthm}

\newfont{\go}{ygoth.tfm scaled 1200}  

\newtheorem{theorem}{Theorem}
\newtheorem{lemma}[theorem]{Lemma}

\newcommand{\cwedge}[1]{\mathop{\wedge}_{{}^{#1}} }
\newcommand{\hook}{\raisebox{-0.35ex}{\makebox[0.6em][r]
{\scriptsize $-$}}\hspace{-0.15em}\raisebox{0.25ex}{\makebox[0.4em][l]{\tiny
 $|$}}}

\numberwithin{equation}{section}
\newcommand{\eps}{\varepsilon}
\newcommand{\tens}[1]{{\boldsymbol{#1}}}

\newcommand{\pa}{{\tens{\partial}}}

\newcommand{\n}[1]{\label{#1}}
\newcommand{\be}{\begin{equation}}
\newcommand{\ee}{\end{equation}}
\newcommand{\ba}{\begin{eqnarray}}
\newcommand{\ea}{\end{eqnarray}}

\newcommand{\beq}{\begin{equation}}
\newcommand{\eeq}{\end{equation}}
\newcommand{\beqa}{\begin{eqnarray}}
\newcommand{\eeqa}{\end{eqnarray}}

\newcommand{\eq}[1]{(\ref{#1})}

\newcommand{\dt}{\tens{d}^T\!}
\newcommand{\delt}{\tens{\delta}^T\!}
\newcommand{\mD}{{\cal D}}

\newcommand\fverb{\setbox\fverbbox=\hbox\bgroup\verb}
\newcommand\fverbdo{\egroup\medskip\noindent%
			\fbox{\unhbox\fverbbox}\ }
\newcommand\fverbit{\egroup\item[\fbox{\unhbox\fverbbox}]}
\newbox\fverbbox

\title{Generalized hidden symmetries and the Kerr--Sen black~hole}

\preprint{DAMTP-2010-23\\
OCU-PHYS 329}

\author{Tsuyoshi Houri\\
Osaka City University Advanced
Mathematical Institute, 3-3-138 Sugimoto, Sumiyoshi, Osaka 558-8585,
JAPAN\\
E-mail: \email{houri@sci.osaka-cu.ac.jp}
}

\author{David Kubiz\v n\'ak\\
DAMTP, University of Cambridge, Wilberforce Road, Cambridge CB3 0WA, UK\\
E-mail: \email{D.Kubiznak@damtp.cam.ac.uk}
}

\author{Claude M. Warnick\\
DAMTP, University of Cambridge, Wilberforce Road, Cambridge CB3 0WA,
UK\\
\emph{and} Queens' College, Cambridge CB3 9ET, UK \\
E-mail: \email{C.M.Warnick@damtp.cam.ac.uk}
}

\author{Yukinori Yasui\\
Department of Mathematics and Physics, Graduate School of Science, Osaka City University, 3-3-138 Sugimoto, Sumiyoshi, Osaka 558-8585, JAPAN\\
E-mail: \email{yasui@sci.osaka-cu.ac.jp}
}

\abstract{
We elaborate on basic properties of 
generalized Killing--Yano
tensors which naturally extend 
Killing--Yano
symmetry in the presence of skew-symmetric torsion. 
In particular, we discuss their relationship to
Killing tensors and the separability of various field equations. 
We further demonstrate that the Kerr--Sen black hole spacetime of heterotic string theory, as well as its  
generalization to all dimensions, 
possesses a generalized closed conformal Killing--Yano 2-form with respect to a torsion identified with the 3-form occuring naturally in the theory. 
Such a  2-form is responsible for complete integrability of geodesic motion as well as for separability of the scalar and Dirac equations 
in these spacetimes. 
}

\keywords{Killing--Yano tensors, string theory black holes, torsion}

\begin{document} 

\section{Introduction}
Killing--Yano symmetry \cite{Yano:1952} is a fundamental hidden
symmetry which plays a crucial r\^ole in higher-dimensional  
rotating black hole spacetimes with spherical horizon topology
\cite{MyersPerry:1986, GibbonsEtal:2004, ChenEtal:2006cqg}, in the case that the supporting matter
consists of a cosmological constant alone. Such black holes  are
uniquely characterized by the existence of this symmetry
\cite{KubiznakFrolov:2007} and derive from it many remarkable
properties, such as complete integrability of geodesic motion 
\cite{PageEtal:2007, KrtousEtal:2007jhep, HouriEtal:2008a}, separability of the scalar
\cite{FrolovEtal:2007, SergyeyevKrtous:2008}, Dirac \cite{OotaYasui:2008, Wu:2008b} and gravitational perturbations \cite{KunduriEtal:2006,  MurataSoda:2008, OotaYasui:2009},
and the special algebraic type of the Weyl tensor \cite{HamamotoEtal:2007, MasonTaghavi:2008}.  
Unfortunately, the demonstrated uniqueness \cite{HouriEtal:2007,
  KrtousEtal:2008, HouriEtal:2009a} prevents these results being extended for black
holes of more general theories with additional matter content, such as
the black holes of various supergravities or string theory.
These black holes are usually much more complicated and the presence
of matter tends to spoil many of the elegant characteristics of their
vacuum brethren. For example, the
Weyl tensor is no longer guaranteed to be algebraically special. 
On the other hand, one may hope that for at least  some of these black
holes one can  define an appropriate generalization of the Killing--Yano
symmetry and infer some of the black hole properties from it.

One possible generalization is an extension of the Killing--Yano
symmetry in the presence of skew-symmetric torsion. This
generalization was first introduced by Bochner and Yano
\cite{YanoBochner:1953} from the mathematical point of view and
recently rediscovered in \cite{Wu:2009a, KubiznakEtal:2009b, Wu:2009b} as a hidden
symmetry of the Chong--Cvetic--L\"u--Pope rotating black hole of $D=5$
minimal gauged supergravity \cite{ChongEtal:2005b}.
More specifically, it was shown that the Chong--Cvetic--L\"u--Pope
black hole admits a `generalized Killing--Yano tensor' if one
identifies the torsion $3$-form with the dual of the Maxwell field,
$\tens{T}=\tens{*F}/\sqrt{3}$. This identification is rather natural
as no additional field is introduced into the theory and the torsion
is `T-harmonic' due to the Maxwell equations. Moreover, the discovered
generalized Killing--Yano tensor shares almost identical properties
with its vacuum cousin; it gives rise to
all isometries of the spacetime \cite{Kubiznak:2009b} and implies
separability of the Hamilton--Jacobi, Klein--Gordon, and Dirac
equations in this background \cite{DavisEtal:2005, Wu:2009b}. Importantly, it was also shown that the
Chong--Cvetic--L\"u--Pope black hole is the unique solution of minimal
gauged supergravity admitting a generalized Killing--Yano tensor
with T-harmonic torsion \cite{AhmedovAliev:2009b}. 
The relationship between the existence of generalized Killing--Yano symmetries 
and separability of the Dirac equation was investigated in \cite{HouriEtal:2010a}.

These results give rise to the natural question of whether there are
some other physically interesting spacetimes which admit 
Killing--Yano tensors with skew symmetric torsion, or whether the
above example is unique, relying on the simplicity of minimal gauged supergravity. 
It is the purpose of this paper to present a family of spacetimes
admitting generalized Killing--Yano symmetry, and hence to show that such symmetry
is more widely applicable.

It is well known, that pseudo-Riemannian manifolds with skew symmetric
torsion occur naturally in superstring theories, where the torsion may
be  identified (up to a factor) with a 3-form field strength occurring
in the theory \cite{Strominger:1986, Agricola:2006}. Black hole spacetimes of
such a theory are natural candidates to admit generalized
Killing--Yano symmetries. We shall consider an effective field theory
describing the low-energy heterotic string theory and demonstrate that
the generalized Killing--Yano symmetry 
appears naturally for the Kerr--Sen solution \cite{Sen:1992}, as well as for its 
higher-dimensional generalizations found by Chow \cite{Chow:2008}. 
The torsion we  identify in both instances is the 3-form field strength $\tens{H}$.
We shall also demonstrate that this symmetry, in common with the
vacuum and minimal supergravity cases, is responsible for 
the complete integrability of geodesic motion and separability of
suitable scalar and Dirac equations in these spacetimes. 

The paper is organized as follows. The general properties of
Killing--Yano tensors with an arbitrary torsion 3-form are studied and
compared to the vacuum case in Sec.~2. The existence of generalized
Killing--Yano symmetries and the corresponding implications for the
separability of various field equations are demonstrated for the
four-dimensional Kerr-Sen black hole in Sec.~3 and for the `charged
Kerr-NUT' spacetimes in all dimensions in Sec.~4. Sec.~5 is devoted to
conclusions.
%, Appendix A contains a sketch of technical calculations justifying the existence of symmetry operators commuting with the Dirac equation.

\section{Generalized Killing--Yano symmetries}
\subsection{Definition}

Throughout this paper, $M^D$ will be a $D$-dimensional spacetime,
equipped with the metric 
\be\label{1}
\tens{g}=g_{ab}\tens{d}x^{a}\tens{d}x^{b}\, .
\ee
Where we need to distinguish between even and odd dimensions, we will set 
$D=2n+\eps\,,$ with $\varepsilon=0,1$ respectively. Henceforth, 
$\{\tens{X}_a \}$ will be an orthonormal basis for $TM$,
$\tens{g}(\tens{X}_a, \tens{X}_b)=\eta_{ab}$, with dual basis
$\{\tens{e}^a\}$ for $T^*M$ with $\tens{g}(\tens{e}^a,
\tens{e}^b)=\eta^{ab}$. We additionally define
\be
\tens{X}^a = \eta^{ab}\tens{X}_b\,, \qquad \tens{e_a} = \eta_{ab} \tens{e}^b.
\ee
Note that $\tens{X}$ will always be a vector regardless of index
position, while $\tens{e}$ is always a $1$-form. In order to state
some of our formulae succinctly it will be convenient to make use of
the $n$-fold contracted wedge product introduced in a recent paper
\cite{HouriEtal:2010a}. This is defined for any $p$-form $\tens{\alpha}$ and $q$-form $\tens{\beta}$ inductively by
\be
\tens{\alpha}
\cwedge{0} \tens{\beta} =\tens{\alpha}
\wedge \tens{\beta}\,, \qquad \tens{\alpha}
\cwedge{n} \tens{\beta} = \tens{X}^a \hook \tens{\alpha} \cwedge{n-1} \tens{X}_a
\hook \tens{\beta}\,,
\ee
where the `hook' operator $\hook$ corresponds to the inner derivative.\footnote{In components the contracted wedge product takes the following form:
\be
({\alpha}
\cwedge{n} {\beta})_{c_1\dots c_{p+q-2n}}=\frac{(p+q-2n)!}{(p-n)!(q-n)!}\,\alpha^{a_1\dots a_n}{}_{[c_1\dots c_{p-n}}
\beta_{|a_1\dots a_n|c_{p-n+1}\dots c_{p+q-2n}]}\,.\nonumber
\ee
}

We wish to consider a connection $\nabla^T$ which has the same geodesics as the Levi-Civita connection and which preserves the metric:
\be
\nabla^T_{\dot{\gamma}} \dot{\gamma} = 0\,, \qquad \nabla^T_X \tens{g} = 0\,.
\ee
Such a connection has totally anti-symmetric torsion which may be identified with a $3$-form, $\tens{T}$, after lowering indices with the metric.
The connection $\tens{\nabla}^T$ acts on a vector field $\tens{Y}$ as
\be\label{j1}
\nabla^T_X \tens{Y}=\nabla_X \tens{Y}+\frac{1}{2}\, \tens{T}(\tens{X}, \tens{Y},
\tens{X}_a) \tens{X}^a\,,
\ee
where $\tens{X}, \tens{Y}$ are vector fields and $\tens{\nabla}$ is the Levi-Civita connection. As usual, we introduce connection 1-forms $(\tens{\omega}^T)^a_{~b}$ by
\be\label{j2}
\nabla^T_{e_b} \tens{X}_a= (\tens{\omega}^T)^c_{~a}(\tens{X}_b) \,\tens{X}_c\,.
\ee
Comparing \eqref{j1} and \eqref{j2} we have
\beq\label{j3}
\tens{\omega}^T_{ab}=\tens{\omega}_{ab}-\frac{1}{2}T_{abc}\tens{e}^c\,,
\eeq
where $\tens{\omega}_{ab}$ is the Levi-Civita connection 1-form which obeys the Cartan relations
\beq\label{j4}
\tens{\omega}_{ab}=-\tens{\omega}_{ba}\,,\quad \tens{de}^a+\tens{\omega}^a_{~b} \wedge \tens{e}^b=0\,.
\eeq
The 1-form $\tens{\omega}^T_{ab}$ satisfies the Cartan relations with torsion
\beq\label{j4t}
\tens{\omega}^T_{ab}=-\tens{\omega}^T_{ba}\,,\quad \tens{de}^a+
(\tens{\omega}^T)^a_{~b} \wedge \tens{e}^b = \tens{T}^a\,,
\eeq
where $\tens{T}_a(\tens{X},\tens{Y})=\tens{T}(\tens{X}_a, \tens{X}, \tens{Y})$\,.

The connection \eq{j1} induces a connection on forms given by
\be\label{j5}
\nabla^T_X \tens{\Psi}=\nabla_X \tens{\Psi}+\frac{1}{2}(\tens{X}\hook \tens{T})\cwedge{1} \tens{\Psi}\,,
\ee
for a $p-$form $\tens{\Psi}$. We additionally define two differential operators related to the exterior derivative and its dual
\ba
\tens{d}^T \tens{\Psi}\!&=&\!\tens{e}^a \wedge \nabla^T_{e_a} \tens{\Psi}
=\tens{d\Psi}-\tens{T} \cwedge{1}\tens{\Psi}\,,
\quad\ \ \label{j6}\\
\tens{\delta}^T \tens{\Psi}\!&=&\!-\tens{e}_a\hook\nabla^T_{e_a}\tens{\Psi}
=\tens{\delta \Psi}-\frac{1}{2}\tens{T} \cwedge{2} \tens{\Psi}\,.\quad\ \ \label{j7}
\ea
These respectively raise and lower the degree of the form.

{\bf Definition.} A {\em generalized conformal Killing--Yano} (GCKY) tensor $\tens{k}$ \cite{KubiznakEtal:2009b} is a $p$-form satisfying
for any vector field $\tens{X}$
\beq\label{CKY}
\nabla^T_X \tens{k}-\frac{1}{p+1}\tens{X}\hook \tens{d}^T \tens{k}+\frac{1}{D-p+1}
\tens{X}^\flat \wedge \tens{\delta}^T \tens{k}=0\,.
\eeq
In analogy with Killing--Yano tensors defined with respect to the Levi-Civita connection, 
a GCKY tensor $\tens{f}$ obeying
$\tens{\delta}^T \tens{f}=0$ is called a {\em generalized Killing--Yano} (GKY) tensor, and a GCKY $\tens{h}$ obeying
$\tens{d}^T \tens{h}=0$ a {\em generalized closed conformal Killing--Yano} (GCCKY) tensor.\\

\subsection{Basic properties}
\begin{lemma} \label{basprop}
GCKY tensors possess the following basic properties:

\begin{enumerate}
\item A GCKY 1-form is equal to a conformal Killing 1-form.
\item The Hodge star $\tens{*}$ maps GCKY $p$-forms into GCKY $(D-p)$-forms. In particular, the Hodge star of a GCCKY $p$-form is a GKY $(D-p)$-form and vice versa.
\item GCCKY tensors form a (graded) algebra with respect to a wedge product, i.e.,  when $\tens{h}_1$ and $\tens{h}_2$ is a GCCKY $p$-form and $q$-form, respectively, then $\tens{h}_3=\tens{h}_1 \wedge \tens{h}_2$ is a GCCKY $(p+q)$-form.
\item Let $\tens{k}$  be a GCKY $p$-form for a metric $\tens{g}$ and a 
torsion 3-form $\tens{T}$. Then, $\tens{\tilde k}=\Omega^{p+1} \tens{k}$ is a GCKY $p$-form for the metric $\tens{\tilde g}=\Omega^2 \tens{g}$ and the torsion $\tens{\tilde T}=\Omega^2 \tens{T}$.
\item Let $\tens{\xi}$ be a conformal Killing vector, $L_\xi \tens{g}=2f\tens{g}$, for some function $f$, and $\tens{k}$ a GCKY $p$-form with torsion $\tens{T}$, obeying 
$L_\xi \tens{T}=2f\tens{T}$. Then 
 $\tens{\tilde k}=L_\xi \tens{k} -(p+1)f\tens{k}$ is a GCKY $p$-form with $\tens{T}$. 
\end{enumerate}
\end{lemma}
\begin{proof}
The properties 1.-3. were proved in \cite{KubiznakEtal:2009b}. Let us 
prove the remaining two properties.
To prove 4., we note that the left hand side of (\ref{CKY}) may be re-written in the form
\begin{eqnarray}
\nabla_X \tens{k}-\frac{1}{p+1}\tens{X}\hook \tens{d} \tens{k}+\frac{1}{D-p+1}
\tens{X}^\flat \wedge \tens{\delta} \tens{k}  && \nonumber \\
+\frac{1}{2}(\tens{X}\hook \tens{T})\cwedge{1}
\tens{k}+\frac{1}{p+1}\tens{X}\hook ( \tens{T}
\cwedge{1}\tens{k})
+ \frac{1}{2(D-p+1)} \tens{X}^\flat \wedge (\tens{T} \cwedge{2} \tens{k})&=& 0\,.
\end{eqnarray}
The first line is the standard conformal
Killing--Yano operator, which transforms homogeneously under a
conformal transformation $\tens{g} \to \Omega^2 \tens{g}$ provided
$\tens{k}\to\Omega^{p+1} \tens{k}$ (see, e.g.,
\cite{BennCharlton:1997}). Note that a $n$-fold contracted wedge
product introduces a conformal factor of $\Omega^{-2n}$. The remaining
terms may then be seen to transform homogeneously with the correct weight to make (\ref{CKY}) conformally invariant provided $\tens{T} \to \Omega^2 \tens{T}$.

Now, suppose we have a family of diffeomorphisms $\phi_t:M \to M$ with
$t \in (-\epsilon, \epsilon)$ such that $\phi_\cdot :  (-\epsilon,
\epsilon)\times M \to M$ is smooth and $\phi_0 = id_M$. We suppose
further that this family of diffeomorphisms are conformal
transformations of $(M, g, T)$, i.e.,
\begin{equation}
\tens{g} = \left(\phi_t \right)_* \left(\Omega(t, x)^2
  \tens{g}\right), \qquad \tens{T} = \left(\phi_t \right)_* \left(\Omega(t, x)^2 \tens{T}\right)\,,
\end{equation}
for some smooth, non-zero function $\Omega :  (-\epsilon,
\epsilon)\times M \to \mathbb{R}$, where clearly $\Omega(0,
x)=1$. Here $ \left(\phi_t \right)_*:T^{\star n}_{\phi_t(x)}M \to
T^{\star n}_x M$ is the pull-back operator. We may differentiate the first equation with respect to $t$ and evaluate the result at $t=0$ to find
\begin{equation}
0 = \left . 2 \Omega \dot{\Omega}\right |_{t=0}\left(\phi_0\right)_*\tens{g} + \Omega^2 \left . \frac{d}{dt}\left(\phi_t \right)_* \tens{g} \right |_{t=0}
\end{equation}
and hence
\begin{equation}
\mathcal{L}_{\xi} \tens{g} = 2 f \tens{g} \label{xicond1}\,,
\end{equation}
where we define $f(x) = -\dot{\Omega}(0, x)$, and $\xi= (d \phi_t /
dt)_{t=0}$. Similarly, the transformation properties of $\tens{T}$ imply
\begin{equation}
\mathcal{L}_{\xi} \tens{T} = 2 f \tens{T}.\label{xicond2}
\end{equation}

Now suppose that $\tens{k}$ is a GCKY $p$-form of the metric
$\tens{g}$ with torsion $\tens{T}$. By the conformal invariance of \eq{CKY}, $\Omega(t,x)^{p+1} \tens{k}$ is a GCKY $p$-form of the metric $\Omega(t,x)^2\tens{g}$ with torsion $\Omega(t,x)^2\tens{T}$. Pulling this back by the diffeomorphism $\phi_t$, we deduce that
\begin{equation}
\tens{k}(t) = \left(\phi_t \right)_* \left(\Omega(t, x)^{p+1} \tens{h}\right) \label{pbck}
\end{equation}
is a GCKY $p$-form of the metric $\tens{g}$ with torsion $\tens{T}$, \emph{for all values of $t$}. In particular so is $\tilde{\tens{k}} = \dot{\tens{k}}(0)$, as solutions of (\ref{CKY}) form a vector space. Differentiating (\ref{pbck}) with respect to $t$ and setting $t=0$, we find
that
\begin{equation}
\tens{\tilde k}=L_\xi \tens{k} -(p+1)f\tens{k}
\end{equation}
is a GCKY $p$-form.

This supposed that we started with a family of conformal
diffeomorphisms, from which we constructed $\xi$. Locally however, we
may start with $\xi$ obeying \eq{xicond1}, \eq{xicond2}  and find a family of conformal diffeomorphisms
for $t \in (-\epsilon, \epsilon)$ such that $\dot{\phi}|_{t=0} = \xi$.

\end{proof}

\subsection{GCKY forms and Killing tensors}
In this section we prove that generalized (conformal) Killing--Yano tensors of arbitrary rank give rise
to (conformal) Killing tensors. 
Conformal Killing tensors are associated with conserved quantities along null geodesics  
which are of higher order in geodesics' momenta.
Let $\gamma$ be an (affine parametrized) null geodesic with
tangent vector ${l}^a={dx^a}/{ds}\,,$ 
\be\label{l}
l^2=\tens{l}\cdot\tens{l}=0\,,\quad \nabla_l^T \tens{l}=\nabla_l \tens{l}=0\,.
\ee 
The requirement that the quantity 
\be\label{C}
C=K_{a_1\dots a_r} l^{a_1}\dots l^{a_r}\,
\ee
is preserved along $\gamma$, that is $\nabla_l C=0$, is taken as a definition for $\tens{K}$ 
to be a conformal Killing tensor \cite{WalkerPenrose:1970}. 
That is, a {\em conformal Killing tensor} $\tens{K}$ of rank $r$ is a symmetric tensor  which obeys  
\be\label{CKT}
K_{a_1a_2 \ldots a_r}=
K_{(a_1a_2 \ldots a_r)}\,,\quad 
\nabla_{(b} K_{a_1a_2 \ldots a_r)}=g_{(b a_1}\tilde{K}_{a_2 \ldots
a_r)}\, .
\ee
The tensor
$\tilde{\tens{K}}$ is  determined by tracing both sides of equation
\eqref{CKT}. If $\tilde{\tens{K}}$ vanishes, the tensor $\tens{K}$ is 
a {\em Killing tensor} \cite{Stackel:1895}. In this case 
the quantity $C$ is preserved also along timelike (spacelike) geodesics $\gamma$ with 
${u}^a={dx^a}/{d\tau}\,,$
\be
\nabla_u^T \tens{u}=\nabla_u \tens{u}=0\,.
\ee

Let us now show how these objects follow from the existence of GCKY tensors.
For a GKY $p$-form $\tens{k}$ and timelike geodesic $\tens{u}$ let us define a $p-1$ form $\tens{w}_{(k)}$,
\be\label{udotk}
\tens{w}_{(k)}=\tens{u}\hook \tens{k}\,.
\ee 
Using the GKY equation, one can easily show that such a form is `torsion' parallel transported, 
\be
\nabla_u^T\tens{w}_{(k)}=\tens{u}\hook \nabla_u^T\tens{k}=\frac{1}{p+1}
\tens{u}\hook \tens{u}\hook \tens{d}^T\tens{k}=0\,.
\ee 
\begin{lemma}\label{kill}
Let $\tens{h}$ and $\tens{k}$ be two GKY tensors of rank $p$. Then 
\be
K_{ab}=h_{(a |c_1\ldots c_{p-1}|}k_{b)}{}^{c_1\ldots c_{p-1}}
\ee
is a Killing tensor of rank 2.
\end{lemma}
\begin{proof} 
We construct $\tens{w}_{(k)}=\tens{u}\hook \tens{k}\,$ and $\tens{w}_{(h)}=\tens{u}\hook \tens{h}\,,$
which automatically satisfy $\nabla^T_u\tens{w}_{(k)}=0$ and $\nabla^T_u\tens{w}_{(h)}=0$. Hence,
any product of $\tens{w}$'s, $\tens{l}$'s, and the metric $\tens{g}$ is torsion parallel propagated along $\gamma$.
In particular, we find 
\be
\nabla^T_u C=\nabla_u C=0\,,\qquad C=\tens{w}_{(k)}\cdot \tens{w}_{(h)}=u^au^bK_{ab}\,.
\ee
This is of the form \eq{C} and hence $K_{ab}$ is a Killing tensor.
\end{proof}

Now, let us consider a GCKY $p$-form $\tens{k}$ and a null geodesic $\tens{l}$. Then the following $p$-form $\tens{F}_{(k)}$:
\be
\tens{F}_{(k)}=\tens{l}^\flat\wedge (\tens{l}\hook \tens{k})\,,
\ee 
obeys 
\be
\nabla_l^T\tens{F}_{(k)}=\tens{l}^\flat\wedge (\tens{l}\hook \nabla_l^T\tens{k})=
\frac{1}{D-p+1}{}\tens{l}^\flat\wedge [\tens{l}\hook (\tens{l}^\flat \wedge \tens{\delta}^T\tens{k})]=0\,. 
\ee 
\begin{lemma}\label{confkill}
Let $\tens{h}$ and $\tens{k}$ be two GCKY tensors of rank $p$. Then 
\be
Q_{ab}=h_{(a |c_1\ldots c_{p-1}|}k_{b)}{}^{c_1\ldots c_{p-1}}
\ee
is a conformal Killing tensor of rank 2.
\end{lemma}
\begin{proof} 
Let $\tens{F}_{(k)}$ and $\tens{F}_{(h)}$ be `torsion' parallel transported forms along $\gamma$ constructed from $\tens{k}$ and $\tens{h}$ as above, $\tens{F}_{(k)}=\tens{l}^\flat\wedge (\tens{l}\hook \tens{k})\,,\ \tens{F}_{(h)}=\tens{l}^\flat\wedge (\tens{l}\hook \tens{h})\,.$
Then, any product of $\tens{F}$'s, $\tens{l}$'s, and the metric $\tens{g}$ is also torsion parallel propagated along $\gamma$. In particular, this is true for the product
\be
(\tens{F}_{(k)}\cdot \tens{F}_{(h)})_{ab}={F}_{(k)} {}_{a}{}^{c_1\dots c_{p-1}} {F}_{(h)} {}_{bc_1\dots c_{p-1}}=l_al_b C\,,
\ee
where $C=l^al^b Q_{ab}$. Hence we have
\be
\nabla_l^T\bigl[(\tens{F}_{(k)}\cdot \tens{F}_{(h)})_{ab}\bigr]=l_al_b \nabla_l^TC=l_al_b \nabla_lC=0\,.
\ee 
This means that $\nabla_lC=0$, and, comparing with \eqref{C}, we
realize that $Q_{ab}$ is a conformal Killing tensor. 
\end{proof}
{\em Remark.} It is obvious from the proofs that both lemmas are valid for `standard' (conformal) Killing--Yano tensors as well.\\

One might ask whether these results admit a converse, in particular:
can all Killing tensors be written as a product of GKY
tensors with respect to a suitable torsion? 
This problem is already complicated in four dimensions in the absence of torsion where it was first considered by Collinson \cite{Collinson:1976} and finally addressed by Ferrando and S\'aez \cite{FerrandoSaez:2002}.
In that case, algebraic and differential conditions are imposed upon the Killing tensor.
It is obvious that introducing arbitrary torsion adds extra degrees of freedom which may be exploited.
Na\"ive counting arguments
suggest that even with these additional degrees of freedom it is not possible to write every Killing tensor as a product of GKY tensors,  however, we have been unable to
exhibit an explicit counter-example.

Let us finally comment on whether a Killing tensor $K_{ab}$ constructed from a GKY p-form induces symmetries of the scalar wave operator $\Box=g^{ab}\nabla_a\nabla_b$. It was demonstrated by Carter \cite{Carter:1977} that a commutator of the symmetry operator $\hat K=\nabla_aK^{ab}\nabla_b\,$ with the scalar wave operator reads
\be
\bigl[\,\Box, \nabla_aK^{ab}\nabla_b\bigr]=\frac{4}{3}\nabla_a\bigl(K_c^{\ [a}R^{b]c}\bigr)\nabla_b\,.
\ee    
The expression on the r.h.s. automatically vanishes whenever the Killing tensor is a square of a Killing--Yano tensor \cite{CarterMcLenaghan:1979} of arbitrary rank. One can easily show that this is no longer true in the presence of torsion and in general torsion anomalies appear on the r.h.s. In other words, GKY p-forms do not in general produce symmetry operators for the Klein--Gordon equation.

\subsection{Spinning particles and the Dirac equation \label{spinpart}}
A key property of Killing--Yano tensors in the absence of torsion (and
other matter fields) is that they are intimately related to an
enhanced worldline supersymmetry of spinning particles in the
semiclassical approximation \cite{GibbonsEtal:1993}. 
Even more interestingly, such property remains true at the `quantum' level. This  
is reflected by the fact that Killing--Yano tensors     
give rise to symmetry operators for the Dirac equation \cite{BennCharlton:1997, Cariglia:2004} and no anomalies appear in the transition. 
This is no longer true in the presence of matter fields. For example,
for the Maxwell field an anomaly appears already at the spinning particle level and destroys the supersymmetry unless the electromagnetic field obeys some additional restrictions \cite{Tanimoto:1995}.
Similarly, it was recently shown  that, up to an explicit anomaly, 
GCKY tensors correspond to an enhanced worldline supersymmetry of spinning particles \cite{RietdijkvanHolten:1996, DeJongheEtal:1996, KubiznakEtal:2009b}
in the presence of torsion
and provide symmetry operators for the (torsion modified) Dirac operator \cite{HouriEtal:2010a}. Let us briefly recapitulate these results.

Since the torsion field naturally couples to particle's spin, it is
not very surprising that the appropriate Dirac operator picks up a torsion correction. 
It was argued in \cite{HouriEtal:2010a} that in the presence of torsion the natural Dirac operator to consider is 
\be\n{DiracT}
\mD=\gamma^a\nabla_a-\frac{1}{24}T_{abc}\gamma^{abc}\,.
\ee  
It was further shown that given a GCKY tensor $\tens{k}$ and provided that the corresponding anomaly terms\footnote{%
It is only the first condition, $\tens{A}_{(cl)}=0$, which emerges from the classical spinning particle approximation.  Correspondingly, we call $\tens{A}_{(cl)}$ a  
{\em classical anomaly} and  $\tens{A}_{(q)}$ (which appears only at the operator level) a `{\em quantum anomaly}'.
}
\ba
\tens{A}_{(cl)}(\tens{k})\!\!&=&\!\!\frac{\tens{d}(\dt \tens{k})}{p+1}-\frac{\tens{T}\wedge \delt \tens{k}}{D-p+1}-\frac{1}{2}
\tens{dT}\cwedge{1}\tens{k}\,,\label{Acl}\\
\tens{A}_{(q)}(\tens{k})\!\!&=&\!\!\frac{\tens{\delta}(\delt \tens{k})}{D-p+1}-\frac{1}{6(p+1)}\tens{T}\cwedge{3}\dt \tens{k}+\frac{1}{12}\tens{dT}\cwedge{3}\tens{k}\,,\label{Aq}
\ea
vanish, one can construct an operator $L_{k}$ which (on-shell)
commutes with $\mD$, $[\mD,L_k]=0$. Such an operator provides an
on-shell symmetry operator for a massless Dirac equation. When
$\tens{k}$ is in addition $\dt$-closed or $\delt$-coclosed the
operator $L_k$ may be modified to produce off-shell (anti)-commuting
operators $M_k$ or $K_k$. 
\begin{lemma}
Let $\tens{f}$ be a GKY $p$-form for which $\tens{A}_{(cl)}(\tens{f})=0=\tens{A}_{(q)}(\tens{f})$. Then,  
the following operator:
\ba\label{Kf}
K_f\!&=&\!f^a_{\ \,b_1\dots b_{p-1}}\gamma^{b_1\dots b_{p-1}}\nabla_a+\frac{1}{2(p\!+\!1)^2}(df)_{b_1\dots b_{p+1}}\gamma^{b_1\dots b_{p\!+\!1}}
\!+\! \frac{1-p}{8(p+1)}T^a_{\ \,b_1b_2}f_{ab_3\dots b_{p+1}}\gamma^{b_1\dots b_{p+1}}
\nonumber\\
&&-\frac{p-1}{4}T^{ab}_{\ \ \,b_1}f_{ab b_2\dots b_{p-1}}\gamma^{b_1\dots b_{p-1}}
+ \frac{(p-1)(p-2)}{24}T^{abc}f_{abc b_1\dots b_{p-3}}\gamma^{b_1\dots b_{p-3}}\,
\ea 
graded anti-commutes with the Dirac operator $\mD$, $\{\mD,K_f\}_+\equiv \mD K_f +(-1)^p K_f \mD=0$. In particular, when $p$ is odd, $K_f$ is a symmetry operator for the massive Dirac equation $(\mD+m)\psi=0$. 
\end{lemma}
The first two terms in \eq{Kf} correspond to the symmetry operator for the `classical' Dirac operator  in the absence of torsion
\cite{BennCharlton:1997, Cariglia:2004}.
The third term is a `leading' torsion correction, present already at the classical spinning particle level \cite{KubiznakEtal:2009b}.
The last two terms are `quantum corrections' due to the presence of torsion. 
Similarly, one has
\begin{lemma} \label{GCCKYoplem}
Let $\tens{h}$ be a GCCKY $p$-form for which $\tens{A}_{(cl)}(\tens{h})=0=\tens{A}_{(q)}(\tens{h})$.
Then, the following operator:
\ba\label{Mh}
M_h\!&=&\!h_{b_1\dots b_{p}}\gamma^{ab_1\dots b_{p}}\nabla_a-\frac{p(D-p)}{2(D-p+1)}(\delta h)_{b_1\dots b_{p-1}}\gamma^{b_1\dots b_{p-1}}- \frac{1}{24}T_{b_1b_2 b_3}h_{b_4\dots b_{p+3}}\gamma^{b_1\dots b_{p+3}}\quad\nonumber\\
&& +\frac{p}{4}T^{a}_{\ \,b_1 b_2}h_{ab_3\dots b_{p+1}}\gamma^{b_1\dots b_{p+1}}
+ \frac{p(p-1)(D-p-1)}{8(D-p+1)}T^{ab}_{\ \ \,b_1}h_{abb_2\dots b_{p-1}}\gamma^{b_1\dots b_{p-1}}\,,
\ea 
graded commutes with the Dirac operator $\mD$, $[\mD,M_h]_-\equiv \mD M_h -(-1)^p M_h \mD=0$.  
\end{lemma}
Since GCCKY tensors form an algebra with respect to the wedge product,
it is natural to ask whether the anomalies respect this algebra. It was shown in
\cite{HouriEtal:2010a} that provided the classical anomaly vanishes for
$\tens{h}_1$ and $\tens{h}_2$, both GCCKY tensors, then it also
vanishes for $\tens{h}_1\wedge\tens{h}_2$.

Note that we assert that if the anomalies vanish then $L_{k}$ is a
symmetry operator. In the case that the anomalies do not vanish, it
may still be possible to modify $L_{k}$ to give a symmetry operator. This is in fact the
case in the $5$-dimensional minimal supergravity case considered in
\cite{KubiznakEtal:2009b}. Although the anomalies do not vanish for
the GKY tensor exhibited, a symmetry operator may nevertheless be
constructed by making use of properties of the torsion in this special
case.

\subsection{GCCKY 2-form}
Let us consider a {\em non-degenerate}\footnote{%
By non-degenerate we mean that the skew symmetric matrix $h_{ab}$ has the maximal possible rank and that its eigenvalues are functionally independent in some spacetime domain. The degenerate case without torsion has been studied in \cite{HouriEtal:2008b, HouriEtal:2009a}
}
 GCCKY 2-form $\tens{h}$
\be\label{PCKY}
\nabla_{X}^T\tens{h}=\tens{X}^{\flat}\wedge \tens{\xi}\,,\qquad  
\tens{\xi}=-\frac{1}{D-1}\tens{\delta}^T\tens{h}\,.
\ee 
In the absence of torsion such an object [called the {\em principal conformal Killing--Yano} (PCKY) tensor]
implies the existence of towers of explicit and hidden symmetries and
determines uniquely (up to $[D/2]$ functions of one variable) the
canonical form of the metric
\cite{KrtousEtal:2007jhep,HouriEtal:2007, KrtousEtal:2008}. In this
subsection we shall see that in the presence of torsion, the GCCKY
2-form $\tens{h}$ is in general a much weaker structure. Our
presentation  closely follows the review \cite{FrolovKubiznak:2008} while we stress some
important differences.

\subsubsection{Canonical basis}
For a non-degenerate $\tens{h}$ one can introduce a Darboux basis in which 
\ba
\tens{g}&=&\delta_{a b}\tens{e}^{a}\tens{e}^{b}=\sum_{\mu=1}^n (\tens{e}^{\mu}\tens{e}^{\mu}+
\tens{e}^{\hat \mu}\tens{e}^{\hat \mu})+\eps \tens{e}^{0}\tens{e}^{0}\, ,\n{gab}\\
\tens{h}&=&\sum_{\mu=1}^n x_{\mu} \tens{e}^{\mu}\wedge \tens{e}^{\hat \mu}\,
,\n{hab}
\ea
where $x_\mu$ are the `eigenvalues' of $\tens{h}$. We refer to $\{\tens{e}\}$ as
the {\em canonical basis} associated with the GCCKY 2-form $\tens{h}$. 
This basis is fixed uniquely up to 2D rotations
in each of the `GKY 2-planes' $\tens{e}^{\mu}\wedge \tens{e}^{\hat \mu}$. This freedom can be exploited, for example,
to simplify the canonical form of the torsion 3-form.

\subsubsection{Towers of hidden symmetries \label{towers}}
According to the property 3 in Lemma \ref{basprop}, the GCCKY 2-form 
generates a tower of GCCKY tensors
\be\label{hj}
\tens{h}^{(j)}\equiv \tens{h}^{\wedge j}=\underbrace{\tens{h}\wedge \ldots \wedge
\tens{h}}_{\mbox{\tiny{total of $j$ factors}}}\, .
\ee 
Because $\tens{h}$ is non-degenerate, one has a
set of $n$ non-vanishing GCCKY $(2j)$-forms $\tens{h}^{(j)}$, $\tens{h}^{(1)}=\tens{h}$. 
In an odd number of spacetime dimensions $\tens{h}^{(n)}$ is dual to a Killing vector $\tens{\eta}=\tens{*{h}}^{(n)}$, whereas in even dimensions  it is proportional to the totally antisymmetric
tensor. 
Contrary to a torsion-less case, forms $\tens{h}^{(j)}$ do not necessary admit a potential, as in general  $(\tens{d}^T)^2\neq 0\neq {\bf d}^T \bf{d}$.
On the other hand, each $\tens{h}^{(j)}$ still gives rise to  a GKY $(D-2j)$-form
\be\label{fj}
\tens{f}^{(j)}\equiv\tens{*}\tens{h}^{(j)}\, ,
\ee
which in its turn generates a Killing tensor $\tens{K}^{(j)}$ by Lemma \ref{kill},
\be\n{Kj}
K^{(j)}_{ab}\equiv{1\over (D-2j-1)!(j!)^2} f^{(j)}_{\, \, \, \, \, a c_1\ldots c_{D-2j-1}}
f_{\, \, \, \, \, b}^{(j) \, \,  c_1\ldots c_{D-2j-1}}\, .
\ee
The choice of the coefficient in the definition \eq{Kj} gives the
Killing tensor an elegant form in the canonical basis [see Eq. \eq{KTdef2} below].
Including the metric $\tens{g}$, which
is a trivial Killing tensor, as the zeroth element, $\tens{K}^{(0)}=\tens{g}$, we obtain a tower of 
$n$ irreducible Killing tensors. 
The explicit form of this tower in the canonical basis is
\ba
\tens{K}^{(j)} \!\!&=&\!\!\sum_{\mu=1}^n A^{(j)}_\mu(\tens{e}^{\mu}\tens{e}^{\mu}+ \tens{e}^{\hat \mu}\tens{e}^{\hat \mu})+\eps A^{(j)}\tens{e}^{0}\tens{e}^{0}\,,\qquad  j=0,\dots, n-1,\label{KTdef2}\\
A^{(j)}\!\!&=&\!\!\sum_{\nu_1<\dots<\nu_j}\! x_{\nu_1}^2\dots x_{\nu_j}^2\;,\qquad
A^{(j)}_\mu=\!\! \sum_{\substack{\nu_1<\dots<\nu_j\\\nu_i\ne\mu}}\!
   x_{\nu_1}^2\dots x_{\nu_j}^2\;.\label{AA}
\ea   
Similar to the torsion-less case one also finds the recursive relation
\be\label{recursive}
\tens{K}^{(j)}=A^{(j)}\tens{g}-\tens{Q}\cdot \tens{K}^{(j-1)}\,,\quad
\tens{K}^{(0)}=\tens{g}\,, 
\ee 
where $\tens{Q}=\tens{h}^2$ is a conformal Killing tensor.
Additionally $\tens{K}^{(i)}\cdot \tens{K}^{(j)}=\tens{K}^{(j)} \cdot
\tens{K}^{(i)}$, which means that $\tens{K}^{(j)}$'s have common
eigenvectors \cite{HouriEtal:2008a}.\footnote{%
As an alternative to the above construction, the tower of Killing tensors $\tens{K}^{(j)}$ can be generated with the help of a generating function  
 $W(\beta)\equiv \det\bigl(I+\sqrt{\beta}w^{-1}F\bigr)\;$~\cite{KrtousEtal:2007jhep, PageEtal:2007}.  Here, $w=u^a u_a$, $\tens{u}$ being the geodesic velocity vector, and $\tens{F}$ is a `torsion-parallel-propagated' 2-form, $\nabla^T_cF_{ab}=0$, which is a projection of $\tens{h}$ along the geodesic, $F_{ab}=P^c_a h_{cd}P^d_b$, 
${P_a^b=\delta_a^b-w^{-1}u^bu_a}$. 
}

A powerful property of $\tens{K}^{(j)}$'s in the absence of torsion is that they  Schouten--Nijenhuis commute
\cite{KrtousEtal:2007jhep, HouriEtal:2008a}. This means that the corresponding integrals of motion for geodesic trajectories (characterized by velocity $\tens{u}$) 
\begin{equation}\label{constfromKT}
  \kappa_j=K^{(j)}_{ab} u^a u^b\; ,
\end{equation}
are {\em in involution}, i.e., they mutually Poisson commute, $\{\kappa_i,\kappa_j\}=0$\,. Contrary to this, in the presence of torsion one rather finds
\begin{equation}\label{PoissonK}
  \bigl[K^{(j)}, K^{(l)}\bigl]_{abc}^T\equiv K^{(j)\,\,e}{}_{(a}\, \nabla_{|e|}^T K^{(l)}_{bc)}- 
  K^{(l)\,\,e}{}_{(a}\, \nabla^T_{|e|} K^{(j)}_{bc)}=0\;,
\end{equation}
which means that $\kappa_j$'s are generally not in involution, unless the torsion $\tens{T}$ obeys some additional conditions.
Similarly, in the absence of torsion the tensors $\tens{K}^{(j)}$ automatically give rise to symmetry operators for the Klein--Gordon equation and forms $\tens{f}^{(j)}$ produce symmetry operators for the Dirac equation. None of these statements remain generally true in the presence of torsion (see previous subsections).
  
The most striking difference between the principal conformal Killing--Yano tensor and a non-degenerate GCCKY 2-form is that the first one generates `naturally' a tower of $n+\eps$ Killing fields whereas the latter does not. More specifically, one can show that in the absence of torsion $\tens{\xi}$, given by \eq{PCKY}, is a (primary) Killing vector and that additional Killing vectors are constructed as 
$\tens{\xi}^{(j)}=\tens{K}^{(j)}\cdot \tens{\xi}$ and $\tens{\eta}$ (in an odd number of dimensions). 
When the torsion is present, neither $\tens{\delta}^T\!\tens{h}$ nor
$\tens{\delta h}$ are in general Killing vectors and the whole
construction breaks down already in the first step; except in odd dimensions one still has at least one Killing field $\tens{\eta}$ derived from $\tens{h}$. 
It is a very interesting open question whether (and if so how) the existence of a non-degenerate GCCKY 2-form $\tens{h}$ implies the existence of $n+\varepsilon$ isometries.
If such construction exists, one can upgrade $n$ natural coordinates $x_\mu$ by adding Killing coordinates to form  a complete {\em canonical basis} as in the case without torsion. Such a result would open a possibility for constructing a `torsion canonical metric'. 
The Kerr--Sen black hole spacetime (and more generally the charged Kerr-NUT metrics) studied in the following two sections provide an example of geometries with a non-degenerate GCCKY 2-form and $n+\varepsilon$ isometries.

\section{Kerr--Sen black hole}

The Kerr--Sen black hole \cite{Sen:1992} is a solution of the low-energy string theory effective action, which in string frame reads
\ba\label{SSen}
S=-\int d^4x\sqrt{-g}e^{-\Phi}\bigl(-R+\frac{1}{12}H_{abc}H^{abc}
-g^{ab}\partial_a\Phi\partial_b\Phi+\frac{1}{8}F_{ab}F^{ab}\bigr)\,.
\ea
Here, $g_{ab}$ stands for the metric in string frame, $\Phi$ is the
dilaton, $\tens{F=dA}$ is the Maxwell field, and there is additionally
a $3$-form
$\tens{H}=\tens{dB}-\frac{1}{4}\tens{A}\wedge\tens{dA}$ where $\tens{B}$ is an antisymmetric tensor field. 
The solution can be obtained by applying the Hassan--Sen transformation \cite{HassanSen:1992} to the Kerr geometry
\cite{Kerr:1963}. 
Its geometry (especially in Einstein frame) is the subject of study of many papers. For example, 
some algebraic properties of this solution were studied in
\cite{Burinskii:1995}, separability of the Hamilton--Jacobi equation in
both (string and Einstein) frames  was proved in \cite{Okai:1994,
  BlagaBlaga:2001}, separability of the charged scalar in Einstein
frame was demonstrated in \cite{WuCai:2003} and the Killing tensor
underlying these results was constructed in
\cite{HiokiMiyamoto:2008}. Although the Einstein frame metric,
$\tens{g}_E=e^{-\Phi}\tens{g}$,  is very similar to the Kerr geometry
and consequently inherits some of its properties, we shall 
see that from the point of view of hidden symmetries it is the string
frame which is more fundamental.\footnote{This will be especially true
  for higher-dimensional generalizations of the Kerr--Sen geometry
  discussed in the next section.}
Namely, we shall demonstrate that the string frame metric $\tens{g}$
possesses a GCCKY 2-form with respect to a natural torsion identified
with the 3-form $\tens{H}$ occurring in the theory. For this reason, in our
study  we mainly concentrate on the string frame.

\subsection{Kerr--Sen black hole in string frame}
\subsubsection{Metric and fields \label{KSmet}}
In Boyer--Lindquist coordinates the string frame Kerr-Sen black hole solution reads \cite{Sen:1992, WuCai:2003}
\ba\label{Senmetric}
ds^2\!&=&\!e^{\Phi}\Bigl\{-\frac{\Delta}{\rho_b^2}\bigl(dt-a\sin^2\!\theta d\varphi\bigr)^2+\frac{\sin^2\!\theta}{\rho_b^2}\Bigl[a dt-(r^2+2br+a^2)d\varphi\Bigr]^2+\frac{\rho_b^2}{\Delta}dr^2+\rho_b^2 d\theta^2\Bigr\}\,,\nonumber\\
\tens{H}\!&=&\!-\frac{2ba}{\rho_b^4}\,\tens{d}t\wedge\tens{d}\varphi\wedge\Bigl[\bigl(r^2-a^2\cos^2\!\theta\bigr)\sin^2\!\theta\tens{d}r-r\Delta\sin 2\theta \tens{d}\theta\Bigr]\,,\nonumber\\
\tens{A}\!&=&\!-\frac{Qr}{\rho_b^2}\bigl(\tens{d}t-a\sin^2\!\theta \tens{d}\varphi\bigr)\,,\nonumber\\
\Phi\!&=&\! 2\ln \left(\frac{\rho}{\rho_b}\right)\,,
\ea
where
\be
\rho^2=r^2+a^2\cos^2\theta\,,\ \quad\  \rho_b^2=\rho^2+2br\,,\ \quad\  \Delta=r^2-2(M-b)r+a^2\,.
\ee
 The solution describes a black hole with mass $M$, charge $Q$,
 angular momentum $J=Ma$, and magnetic dipole momentum $\mu=Qa$. When
 the twist parameter $b=Q^2/2M$ is set to zero, the solution reduces to the Kerr geometry, which can be easily recognized in the brackets.

Let us choose the following basis of 1-forms:
\ba\label{KSbasis}
\tens{e}^{0} \!\!&=&\!\frac{\rho\sqrt{\Delta}}{\rho_b^2}\,\bigl(\tens{d}t-a\sin^2\!\theta\tens{d}\varphi\bigr)\,,\quad
\tens{e}^{1}=\frac{\rho}{\sqrt{\Delta}}\,\tens{d}r\,,\nonumber\\
\tens{e}^{2} \!\!&=&\!\!\frac{\rho \sin\!\theta}{\rho_b^2}\Bigl[a dt-(r^2+2br+a^2)d\varphi\Bigr]\,,\quad 
\tens{e}^{3}=\rho\tens{d}\theta\,,\quad\ \ 
\ea
and define the functions 
\be
T_{0}^{\pm}=-\frac{2a\sin\theta}{\rho}\left(-\frac{r+b}{\rho_b^2}\pm\frac{r}{\rho^2}\right)\,,\quad  
T_{1}^{\pm}=-\frac{2a\cos\theta\sqrt{\Delta}}{\rho}\left(-\frac{1}{\rho_b^2}\pm\frac{1}{\rho^2}\right)\,.
\ee
Then, the metric and the fields take the form
\be\label{fields}
\tens{g}=-{\tens{e}^0}^2+{\tens{e}^1}^2+{\tens{e}^2}^2+{\tens{e}^3}^2\,,\quad
\tens{H}=T_0^+\tens{e}^{012}+T_1^+\tens{e}^{023}\,,\quad 
\tens{A}=-\frac{Qr}{\rho \sqrt{\Delta}}\tens{e}^0\,,
\ee
and the inverse metric is given by 
\ba\label{X}
\tens{X}_{0} \!\!&=&\!\frac{1}{\rho\sqrt{\Delta}}\,
\Bigl[(r^2+2br+a^2)\pa_t+a\pa_\varphi\Bigr]\,,\quad
\tens{X}_{1}=\frac{\sqrt{\Delta}}{\rho}\,\pa_r\,,\nonumber\\
\tens{X}_{2} \!\!&=&\!\!-\frac{1}{\rho \sin\theta}\,(a\sin^2\!\theta\pa_t+\pa_\varphi)\,,\quad
\tens{X}_{3}=\frac{1}{\rho}\,\pa_\theta\,.
\ea
In order to separate the Dirac equation, we shall also need the spin connection. This can be obtained from the Cartan's equation 
$\tens{de}^a+\tens{e}^a_{\ b}\wedge \tens{e}^b=0$ and is given as follows: 
\ba\label{spcon}
\tens{\omega}_{01}\!&=&\!-A\tens{e}^0-B\tens{e}^2\,,\quad 
\tens{\omega}_{02}=-B\tens{e}^1+C\tens{e}^3\,,\quad 
\tens{\omega}_{03}=-D\tens{e}^0-C\tens{e}^2\,,\nonumber\\
\tens{\omega}_{12}\!&=&B\tens{e}^0
-E\tens{e}^2\,,\quad
\tens{\omega}_{13}=D\tens{e}^1-E\tens{e}^3\,,\quad
\tens{\omega}_{23}=-C\tens{e}^0-F\tens{e}^2\,,
\ea
where
\ba
A&=&\frac{\rho_b^2}{\rho^2}\frac{d}{dr}\Bigl(\frac{\rho\sqrt{\Delta}}{\rho_b^2}\Bigr)\,,\quad
B=\frac{a(r+b)\sin\theta}{\rho\rho_b^2}\,, \quad
C=\frac{a\cos\theta\sqrt{\Delta}}{\rho\rho_b^2}\,,\nonumber\\
D&=&-\frac{a^2\sin\theta\cos\theta}{\rho^3}\,,\quad 
E=\frac{r\sqrt{\Delta}}{\rho^3}\,,\quad 
F=-\frac{1}{\sin\theta}\frac{\rho_b^2}{\rho^2}\frac{d}{d\theta}\Bigl(\frac{\rho\sin\theta}{\rho_b^2}\Bigr)\,.
\ea

\subsubsection{Hidden symmetries}
Besides two obvious isometries $\pa_t$ and $\pa_\varphi$, the Kerr--Sen geometry admits an irreducible Killing tensor
\be\label{KTSen}
\tens{K}=a^2\cos^2\!\theta(\tens{e}^{0}\tens{e}^{0}-\tens{e}^{1}\tens{e}^{1})+
r^2(\tens{e}^{2}\tens{e}^{2}+\tens{e}^{3}\tens{e}^{3})\,.
\ee
Such a tensor is responsible for separability of (charged)
Hamilton--Jacobi equation and hence for the complete integrability of
the motion of (charged) particles.  Moreover, the metric
possesses two GCCKY 2-forms. The first one naturally generalizes the
closed conformal Killing--Yano 2-form of the Kerr geometry with respect to the
torsion identified with 3-form $\tens{H}$.  More specifically, if we
identify 
\be
\tens{T}_+=\tens{H}\,,
\ee
where the 3-form $\tens{H}$ is given by \eq{fields}, then 
one can explicitly verify that 
\be
\tens{h}_+=r\tens{e}^0\wedge\tens{e}^1+a\cos\theta\tens{e}^2\wedge\tens{e}^3\,
\ee
is a GCCKY 2-form obeying \eq{PCKY}. 
This is true with no restriction on the function $\Delta=\Delta(r)$.
Contrary to the Kerr case, one does not simply recover the isometries
from the divergence of $\tens{h}_+$; one has 
\be
\tens{\xi}_+=-\frac{1}{3}\tens{\delta}^{T_+} (\tens{h}_+)=-\frac{\sqrt{\Delta}}{\rho}\,\tens{e}^0+\frac{a\sin\theta}{\rho}\, \tens{e}^2\,, \qquad \tens{\xi}_+^\sharp=e^{-\Phi}\pa_t\,.
\ee 
One can easily check that in the limit $b=0$, the torsion $\tens{T}_+$ vanishes and one recovers
the standard form of the Kerr geometry and its corresponding PCKY tensor, as found by Floyd and Penrose \cite{Floyd:1973, Penrose:1973}.
Let us also observe that if one identifies $x_1=r$ and
$x_2=a\cos\theta$, one recovers a canonical basis for
$\tens{h}_+$. This leads to a transformation to the `canonical
coordinates' and the Carter--Plebanski-like form \cite{Carter:1968pl, Plebanski:1975} of the
Kerr--Sen geometry (see also Section \ref{CPSec}).

There is yet another GCCKY 2-form in the Kerr--Sen geometry,
\be
\tens{h}_-=r\tens{e}^0\wedge\tens{e}^1-a\cos\theta\tens{e}^2\wedge\tens{e}^3\,,\qquad
\tens{\xi}_-=-\frac{1}{3}\tens{\delta}^{T_-} (\tens{h}_-)=-\frac{\sqrt{\Delta}}{\rho}\,\tens{e}^0-\frac{a\sin\theta}{\rho}\, \tens{e}^2\,,
\ee
with respect to a different torsion $\tens{T}_-$ given by
\be
\tens{T}_-=T_{0}^-\tens{e}^{012}+T_{1}^-\tens{e}^{023}\,.   
\ee
Such a torsion is rather peculiar.  It remains non-trivial 
in the limit of the Kerr geometry where one has 
\be
\tens{T}\equiv (\tens{T}_-)|_{b=0}=\frac{4a}{\rho^3}\Bigl(r\sin\theta \tens{e}^{012}+\cos\theta\sqrt{\Delta}\tens{e}^{023}\Bigr)|_{b=0}\,,
\ee
and
\be
\tens{\delta T}=0\quad \Leftrightarrow \quad  \tens{T}=\tens{*d}\alpha\,,\quad \alpha=4\arctan\Bigl( \frac{a\cos\theta}{r}\Bigr)\,.
\ee
Although such a torsion seems unfamiliar and cannot be related to
fields occurring naturally in the theory, it is `encoded' in the
geometry of the  spacetime. Whether it is of some physical interest
remains an open question.

Let us finally remark that contrary to the GCCKY tensor in the
Chong--Cvetic--L\"u--Pope black hole spacetime of minimal supergravity
\cite{Wu:2009a, KubiznakEtal:2009b, Wu:2009b},   neither the 2-form $\tens{h}_+$
nor $\tens{h}_-$ are closed and hence neither can be generated from a potential.
We shall see in Section \ref{KSDir} that both these 2-forms give rise
to symmetry operators of appropriately modified Dirac equations in the
Kerr--Sen black hole background. 

\subsubsection{Motion of charged particles}
Motion  of test particles in the string frame of the Kerr--Sen black
hole background is studied in detail in
\cite{BlagaBlaga:2001}; it is completely integrable due to the Killing
tensor \eq{KTSen}. Let us here, for completeness, demonstrate that the
same remains true also for charged particles.  The motion of a
particle with charge $e$ is governed by the minimally coupled
Hamilton--Jacobi equation   
\be
\frac{\partial S}{\partial \lambda}+g^{ab}(\partial_a S+eA_a)(\partial_bS+eA_b)=0\,.
\ee 
Using \eq{X} and \eq{fields}, this equation takes the following explicit form
\ba
\frac{\partial S}{\partial \lambda}-\frac{1}{\rho^2\Delta}\,
\Bigl[(r^2+2br&+&a^2)\partial_t S+a\partial_\varphi S-eQr\Bigr]^2
+\frac{{\Delta}}{\rho^2}\,(\partial_r S)^2\,\nonumber\\
&+&\frac{1}{\rho^2 \sin^2\!\theta}\,(a\sin^2\!\theta\partial_t S+\partial_\varphi S)^2
+\frac{1}{\rho^2}\,(\partial_\theta S)^2=0\,.
\ea 
It allows a separation of variables
$S=-\lambda\kappa_0-Et+L\varphi+R(r)+\Theta(\theta)$, where the
functions $R(r)$ and $\Theta(\theta)$ obey the ordinary differential equations
\ba
R'^2-\frac{W_r^2}{\Delta^2}-\frac{V_r}{\Delta}&=&0\,,\qquad W_r=-E(r^2+2br+a^2)+aL-eQr\,,\ \ V_r=\kappa +\kappa_0 r^2\,,\nonumber\\
\Theta'^2+\frac{W_\theta^2}{\sin^2\!\theta}-V_\theta&=&0\,,\qquad W_\theta=-aE\sin^2\!\theta+L\,,\ \  
V_\theta=-\kappa+\kappa_0 a^2\cos^2\!\theta.
\ea 
Identifying $p_a=\partial_aS+eA_a$, we find the particle's momentum $\tens{p}$ (obeying $\dot p_a=-eF_{ab}p^b$)  
\be
\tens{p}=-\Bigl(E+\frac{eQr}{\rho_b^2}\Bigr)\tens{d}t+\Bigl(L+\frac{aeQr\sin^2\!\theta}{\rho_b^2}\Bigr)\tens{d}\varphi+
\sigma_r\sqrt{\frac{W_r^2}{\Delta^2}\!+\!\frac{V_r}{\Delta}}\,\tens{d}r+\sigma_\theta\sqrt{V_\theta\!-\!\frac{W_\theta^2}{\sin^2\!\theta}}\,\tens{d}\theta\,.
\ee
Here, $\sigma_r, \sigma_\theta=\pm$ are independent signs, parameters
$E$ and $L$ are separation constants corresponding to the Killing
fields $\pa_t$ and $\pa_\varphi$, $\kappa_0$ is the normalization of
the momentum, and $\kappa$ denotes a separation constant associated
with the Killing tensor \eq{KTSen}.

\subsubsection{Separability of the charged scalar field equation}

Given that the Hamilton--Jacobi equation for the motion of charged
particles separates for the Sen black hole in string frame, it is
natural to consider whether the equations for a charged scalar field
separate. The matter content of the theory is determined by
\eq{SSen}, and contains a scalar field in the form of the dilaton
$\Phi$. In order to remain within this model, one should consider
perturbations of the fields appearing in the action. Since
all the fields have non-trivial background values, the linear
perturbations of the fields couple to one another, so that one may not consistently
consider a linearised perturbation of one field in isolation. In order
to circumvent this problem, we introduce a new charged scalar field
which vanishes in the background and consider perturbations of
this. One might hope that analysing such a test field may give some
insight into the dynamics of perturbations in the background, while
remaining tractable.

A reasonable guess for the appropriate field equation of a charged
scalar field in this background would be the minimally coupled
Klein--Gordon equation. This does not separate in string
frame. Considering the action \eq{SSen}, one concludes that
the na\"ive Klein--Gordon equation is not the most natural equation for
a charged scalar field in this background. We instead consider a field
$\psi$ whose equations of motion derive from the action
\be
S = \int d^4x\sqrt{-g}\,\frac{e^{-\Phi}}{2}\,\bigl(g^{ab} \mathcal{D}_a \psi
\mathcal{D}_b \psi +m^2 \psi^2\bigr)\,.
\ee
We introduce here the gauge covariant derivative $\mathcal{D}_a
= \nabla_a +ieA_a$, where $\tens{A}$ is the background $1$-form
field given in Section \ref{KSmet}. In the case $m=0$, this field obeys the standard
charged massless Klein--Gordon equation when we transform to Einstein
frame. In that frame the charged Klein--Gordon equation has been
separated by Wu and Cai \cite{WuCai:2003}.

For the particular $\tens{A}$, $\Phi$ of the Kerr-Sen background, we find that
$\nabla_a A^a=A^a \nabla_a \Phi = 0$, so the equations which arise upon
varying $S$ may be written as
\begin{equation}
\nabla^a\nabla_a \psi - \nabla^a \Phi \nabla_a \psi + 2 i e A^a
\nabla_a \psi + e^2 A^2\psi - m^2 \psi = 0. \label{chgscal}
\end{equation}
A short calculation with the line element given above shows that
$\sqrt{-g} = (\rho^4/\rho_b^2) \sin \theta$. Making use of the
expression $\nabla_a \nabla^a \psi = (-g)^{-1/2} \partial_a
(\sqrt{-g}g^{ab} \partial_b \psi)$ we find that \eq{chgscal}
separates multiplicatively with the ansatz
\be
\psi = R(r) \Theta(\theta) e^{-i \omega t + i h \varphi}\,.
\ee
The resulting ordinary differential equations for the functions $R(r),
\Theta(\theta)$ are
\ba
\frac{1}{R} \frac{d}{dr} \left(\Delta \frac{dR}{dr} \right)+
\frac{U_r^2}{\Delta} - r^2 m^2-\kappa\!&=&\!0\,,\nonumber\\
\frac{1}{\Theta}\frac{1}{ \sin \theta}\frac{d}{d\theta}
\left(\sin\theta \frac{d\Theta}{d\theta} \right)-
\frac{U_\theta^2}{\sin^2\theta} - a^2 m^2 \cos^2 \theta-\kappa \!&=&\!0\,,\qquad\qquad 
\ea
where the potential functions $U_r, U_\theta$ are given by 
\be
U_r = ah -\omega(r^2+2br+a^2)-eQr, \qquad
U_\theta = h-a\omega \sin^2 \theta, \label{Upot}
\ee
and $\kappa$ is a separation constant, related to the Killing tensor \eq{KTSen}.

\subsubsection{Separability of the Dirac equation \label{KSDir}}

The torsion modified Dirac equation for a particle carrying charge $e$ reads
\be\label{DiracSen}
\bigl[\gamma^a(D_\pm)_a+m\bigr]\psi_\pm=0\,,\quad\  (D_\pm)_a=X_a+\frac{1}{4}\gamma^b\gamma^c\omega_{bc}(X_a)-\frac{1}{24}\gamma^b\gamma^c(T_\pm)_{abc}+ieA_a\,.
\ee
Using the connection \eq{spcon} and the inverse basis \eq{X} we find its explicit form  
\ba\label{Diracexplicit}
\Bigl\{\!\!&&\frac{\gamma^0}{\rho\sqrt{\Delta}}\Bigl[(r^2+2br+a^2)\partial_t+a\partial_\varphi-ieQr \Bigr]+\gamma^1\Bigl(E+\frac{A}{2}+\frac{\sqrt{\Delta}}{\rho}\partial_r\Bigr)-\frac{\gamma^2}{\rho\sin\theta}\bigl(a\sin^2\!\theta\partial_t
+\partial_\varphi\bigr)\nonumber\\
&&+\gamma^3 \Bigl(D-\frac{F}{2}+\frac{1}{\rho}\partial_\theta\Bigr)+\frac{\gamma^{012}}{4}(2B-T_0^\pm)+\frac{\gamma^{023}}{4}(2C-T_1^\pm)+ m\Bigr\}\psi_\pm=0\,.
\ea
We use the following representation of gamma matrices $\{\gamma^a,\gamma^b\}=2\eta^{ab}$:
\be
\gamma^0=\left(   
\begin{array}{cc}
0 & -I\\
I& 0
\end{array}
\right)\, ,\quad
\gamma^1=\left(   
\begin{array}{cc}
\ 0 & \ I\\
\ I& \ 0
\end{array}
\right)\,, \quad
\gamma^2=\left(   
\begin{array}{cc}
\sigma^2 & 0\\
0& -\sigma^2
\end{array}
\right)\, ,\quad
\gamma^3=\left(   
\begin{array}{cc}
\sigma^1 & 0\\
0& -\sigma^1
\end{array}
\right)\, ,
\ee
where $\sigma^i$ are Pauli matrices. 

To keep track of various signs, we first consider the case of $\tens{T}_+$.
Separation of the Dirac equation can be achieved with the ansatz 
\be
\psi_+=\frac{\rho_b}{\rho}\left(   
\begin{array}{c}
(r+ia\cos\theta)^{-1/2}R_+S_+\\
(r-ia\cos\theta)^{-1/2}R_+S_-\\
(r-ia\cos\theta)^{-1/2}R_-S_+\\
(r+ia\cos\theta)^{-1/2}R_-S_-
\end{array}
\right)\,e^{i(h\varphi-\omega t)}\,,
\ee
with functions $R_{\pm}=R_{\pm}(r)$ and $S_{\pm}=S_{\pm}(\theta)$. 
Inserting this ansatz in \eq{Diracexplicit}, we obtain eight equations with four separation constants.
The consistency of these equations implies that only one of the separation constants is independent, we denote it by $\kappa$. Finally one obtains the following 
four coupled first order ordinary differential equations for $R_{\pm}$ and $S_{\pm}$:
\ba
\frac{d R_\pm}{dr}+R_\pm\frac{\Delta'\pm 4iU_r}{4\Delta}+R_{\mp}\frac{mr\mp \kappa}{\sqrt{\Delta}}&=&0\,,\nonumber\\
\frac{d S_\pm}{d\theta}+S_\pm\frac{\cos\theta\pm 2U_\theta}{2\sin\theta}+S_{\mp}\bigl(\pm ima\cos\theta-\kappa\bigr)&=&0\,,
\ea
where $U_r$ and $U_\theta$ are given by Eq. \eq{Upot} of the previous section.

Similarly, for $\tens{T}_-$ one has the separation ansatz 
\be
\psi_-=\frac{\rho_b}{\rho}\left(   
\begin{array}{c}
(r-ia\cos\theta)^{-1/2}R_+S_+\\
(r+ia\cos\theta)^{-1/2}R_+S_-\\
(r+ia\cos\theta)^{-1/2}R_-S_+\\
(r-ia\cos\theta)^{-1/2}R_-S_-
\end{array}
\right)\,e^{i(h\varphi-\omega t)}\,,
\ee
and the functions $R_{\pm}$ and $S_{\pm}$ satisfy the following coupled ODEs:
\ba
\frac{d R_\pm}{dr}+R_\pm\frac{\Delta'\pm 4iU_r}{4\Delta}+R_{\mp}\frac{mr\mp \kappa}{\sqrt{\Delta}}&=&0\,,\nonumber\\
\frac{d S_\pm}{d\theta}+S_\pm\frac{\cos\theta\pm 2U_\theta}{2\sin\theta}+S_{\mp}\bigl(\mp ima\cos\theta-\kappa\bigr)&=&0\,.
\ea
This result is new and non-trivial even in the limit where the Kerr
geometry is recovered. 

In both cases, separability can be justified by general theory of
Section \ref{spinpart}. 
Indeed, one can easily show that the anomalies \eq{Acl} and \eq{Aq} reduce to
\be
(\tens{A}_\pm)_{(cl)}=\tens{T}_\pm\wedge \tens{\xi}_\pm\,,\quad
(\tens{A}_\pm)_{(q)}=-\tens{\delta}(\tens{\xi}_\pm)\,,
\ee
and in both considered cases these vanish.  The corresponding symmetry operators that commute with the Dirac operator \eq{DiracSen} are therefore given by Lemma \ref{GCCKYoplem} and read 
\be\label{Mpm}
M_\pm=(h_{\pm})_{bc}\gamma^{abc}\nabla_a-\frac{2}{3}(\delta h_\pm)_{a}\gamma^{a}+\frac{1}{2}(T_\pm)^a{}_{bc}(h_{\pm})_{ad}\gamma^{bcd}+\frac{1}{12}(T_\pm)^{ab}{}_c(h_{\pm})_{ab}\gamma^c\,.
\ee  
It can be explicitly verified that these commute with the Dirac
operator.  As expected, the demonstrated separability is underpinned by
the existence of GCCKY tensors.

\subsection{Kerr--Sen black hole in Einstein frame}
Let us now briefly consider the Kerr--Sen geometry in Einstein frame, $\tens{g}_E=e^{-\Phi}\tens{g}$. Introducing an ortonormal basis of 1-forms 
\ba\label{KSbasisE}
\tens{e}^{0}_E \!&=&\!\frac{\sqrt{\Delta}}{\rho_b}\,\bigl(\tens{d}t-a\sin^2\!\theta\tens{d}\varphi\bigr)\,,\quad
\tens{e}^{1}_E=\frac{\rho_b}{\sqrt{\Delta}}\,\tens{d}r\,,\nonumber\\
\tens{e}^{2}_E \!&=&\!\frac{\sin\theta}{\rho_b}\Bigl[a dt-(r^2+2br+a^2)d\varphi\Bigr]\,,\quad 
\tens{e}^{3}_E=\rho_b\tens{d}\theta\,, 
\ea
the metric reads
\be\label{SengE}
\tens{g}_E=-{\tens{e}^0_E}^2+{\tens{e}^1_E}^2+{\tens{e}^2_E}^2+{\tens{e}^3_E}^2\,.
\ee
Obviously, the Kerr--Sen metric in Einstein frame is very similar to the Kerr geometry. Consequently it shares some of its miraculous properties. In particular, this is true 
for separability of various field equations in its background and for the existence of hidden symmetries.
Let us first concentrate on hidden symmetries.

The metric possesses an irreducible Killing tensor \cite{HiokiMiyamoto:2008}
\be\label{K2}
\tens{K}=a^2\cos^2\!\theta\bigl(\tens{e}^{0}_E\tens{e}^{0}_E-
\tens{e}^{1}_E\tens{e}^{1}_E\bigr)
+r(r+2b)\bigl(\tens{e}^{2}_E\tens{e}^{2}_E+
\tens{e}^{3}_E\tens{e}^{3}_E\bigr)\,,
\ee
which is responsible for complete integrability of motion of charged particles and separability of the charged scalar field \cite{WuCai:2003}.
Moreover, the metric admits two GCCKY 2-forms 
\be
\tens{h}_\pm=\sqrt{r(r+2b)}\tens{e}^0_E\wedge\tens{e}^1_E\pm a\cos\theta\tens{e}^2_E\wedge\tens{e}^3_E\,
\ee
with respect to the torsions
\be
\tens{T}_\pm=T_0^\pm\tens{e}_E^{012}+T_1^\pm\tens{e}_E^{023}\,,
\ee
where 
\be
T_{0}^{\pm}=\frac{2a\sin\theta}{\rho_b^3}\left[r+b\mp \sqrt{r(r+2b)}\right]\,,\quad  
T_{1}^{\pm}=\mp\frac{2a\cos\theta\sqrt{\Delta}}{\rho_b^3 \sqrt{r(r+2b)}}\left(r+b\mp \sqrt{r(r+2b)}\right)\,.
\ee
One can easily show that for both choices, anomalies   \eq{Acl} and \eq{Aq} vanish and operators \eq{Mpm} give commuting operators for the corresponding Dirac equations. 
By calculations analogous to Section \ref{KSDir}, one can demonstrate that both Dirac equations separate. The meaning of torsions $\tens{T}_\pm$ is at the moment unclear. 
Let us finally mention that as is the case in the Kerr geometry
\cite{FrolovEtal:1989}, one can prove complete integrability of stationary
string configurations in the background \eq{SengE}.

\section{Higher-dimensional charged Kerr-NUT spacetimes}
Let us now consider the following higher-dimensional generalization of the string frame action \eq{SSen}:
\be\label{S}
S=\int_{M^D} e^{\phi\sqrt{D/2-1}}\Bigl(\tens{*}R+\frac{D-2}{2}\,\tens{*d}\phi\wedge \tens{d}\phi
   -\tens{*F}\wedge \tens{F}-\frac{1}{2}\tens{*H}\wedge \tens{H}\Bigr) \,,
\ee
where $\tens{F}=\tens{dA}$ and $\tens{H}=\tens{dB}-\tens{A}\wedge \tens{dA}$.
The system consists of a metric $g_{ab}$, scalar field $\phi$, 
$U(1)$ potential $\tens{A}$, and 2-form potential $\tens{B}$.
This kind of action gives a bosonic part of supergravity 
such as heterotic supergravity compactified on a torus.
The corresponding equations of motion are 
\begin{align}
& R_{ab}-\frac{1}{2}R g_{ab}= 
  \sqrt{\frac{D-2}{2}}\,\nabla_a\nabla_b\phi 
  -\sqrt{\frac{D-2}{2}}\,g_{ab}\nabla^2\phi-\frac{D-2}{4}\,g_{ab}\bigl(\nabla\phi\bigr)^2 \nonumber\\
& \hspace{2.0cm} 
+\Big(F_a{}^cF_{bc}-\frac{1}{4}g_{ab}F^2\Big)
  +\frac{1}{4}\Big(H_a{}^{cd} H_{bcd}-\frac{1}{6}g_{ab}H^2\Big)\,, \nonumber\\
& \tens{d}\Big(e^{\phi\sqrt{D/2-1}}\tens{*F}\Big)+(-1)^{D}e^{\phi\sqrt{D/2-1}}\tens{*H}\wedge \tens{F}=0\,,\quad 
  \tens{d}\Big(e^{\phi\sqrt{D/2-1}}\tens{*H}\Big)=0 ~, \nonumber\\
& R-\frac{D-2}{2}\bigl(\nabla\phi\bigr)^2
  -\sqrt{2(D-2)}\nabla^2\phi
  -\frac{1}{2}F^2-\frac{1}{12}H^2 = 0 ~. \label{eomchow}
\end{align}
Alternatively, one could consider an Einstein frame metric,
$\tens{g}_E=e^{\phi\sqrt{2/(D-2)}}\tens{g}$. After this
transformation, the action \eq{S} is equivalent to the action (2.1) in
\cite{Chow:2008} in the case when two scalar and two $U(1)$ charges
are set equal.

\subsection{Metric and fields \label{CPSec}}
The `charged Kerr-NUT' solution of the theory \eq{S} in all dimensions was obtained by Chow in \cite{Chow:2008}. It generalizes the Kerr--Sen solution for all $D$ as well as the Kerr-NUT solution of Chen, L\"u, and Pope \cite{ChenEtal:2006cqg} by including matter fields $\phi$, $\tens{A}$ and $\tens{B}$. 
The metric and the fields are given by\footnote{%
As usual for these kind of solutions (see, e.g.,
\cite{ChenEtal:2006cqg, KubiznakFrolov:2007}), we work with an
analytically continued metric, where one of the $x_\mu$ correspond to
the Wick rotated radial coordinate $r$ and (in even dimensions) the
corresponding parameter $m_\mu$  is imaginary mass. The advantage of
this continuation is that the radial and longitudinal coordinates
appear on exactly the same footing and the metric takes an extremely
symmetric form. Let us stress that working in this continuation
affects neither the existence of hidden symmetries nor separability of
the field equations studied below.
}
\begin{align}
&
\tens{g}= \sum_{\mu=1}^n\frac{\tens{d}x_\mu^2}{Q_\mu}
      +\sum_{\mu=1}^nQ_\mu\Big(\tens{{\cal A}}_\mu-\sum_{\nu=1}^n\frac{2N_\nu s^2}{HU_\nu}\,\tens{{\cal A}}_\nu\Big)^2
      +\varepsilon S\Big(\tens{{\cal A}}-\sum_{\nu=1}^n\frac{2N_\nu s^2}{HU_\nu}\,\tens{{\cal A}}_\nu\Big)^2\,, \nonumber\\
&
\phi=\sqrt{\frac{2}{D-2}}\ln H\,,\quad \tens{A}=\sum_{\nu=1}^n\frac{2N_\nu sc}{HU_\nu}\tens{{\cal A}}_\nu ~,~~ \nonumber\\
&
\tens{B}=\Big(\sum_{k=0}^{n-1}(-1)^kc_{n-k-1} \tens{d}\psi_k+\varepsilon\,\tilde{c}\tens{d}\psi_n\Big)\wedge 
\Big(\sum_{\nu=1}^n\frac{2N_\nu s^2}{HU_\nu}\tens{{\cal A}}_\nu\Big) ~, \label{Chow solution}
\end{align}
where we have defined the following 1-forms:
\begin{align}
\tens{{\cal A}}_\mu = \sum_{k=0}^{n-1}A_\mu^{(k)}\tens{d}\psi_k\,,\qquad \tens{{\cal A}}=\sum_{k=0}^n A^{(k)}\tens{d}\psi_k ~,
\end{align}
and the following functions:
\begin{align}
&
H = 1+\sum_{\mu=1}^n\frac{2N_\mu s^2}{U_\mu} ~,~~ N_\mu = m_\mu x_\mu^{1-\varepsilon} ~,~~
S=\frac{\tilde{c}}{A^{(n)}} ~, \nonumber\\
&
Q_\mu = \frac{X_\mu}{U_\mu} ~,~~
U_\mu = \prod_{\substack{\nu=1\\ \nu\neq\mu}}^n (x_\mu^2-x_\nu^2) ~,~~
X_\mu = \sum_{k=0}^{n-1}c_k x_\mu^{2k}+2N_\mu+\varepsilon\,\frac{(-1)^n\tilde{c}}{x_\mu^2} ~, \\
&
A_\mu^{(k)} = \sum_{\substack{1\leq \nu_1<\dots<\nu_k\leq n\\ \nu_i\neq\mu}}x_{\nu_1}^2\cdots x_{\nu_k}^2 ~,~~
A^{(k)} = \sum_{1\leq \nu_1<\dots<\nu_k\leq n}x_{\nu_1}^2\cdots x_{\nu_k}^2 ~,~~ 
A_\mu^{(0)}=A^{(0)}=1 ~. \nonumber
\end{align}
We have introduced $s=\sinh\delta$, $c=\cosh\delta$, $c_{n-1}=-1$, and
$m_\mu$ ($\mu=1,\dots,n$), $c_k$ ($k=0,\dots,n-2$), $\tilde{c}$, and $\delta$ are arbitrary constants. We have verified directly that these fields satisfy the equations of motion \eq{eomchow}.

Let us remark that the Kerr-NUT solution \cite{ChenEtal:2006cqg} is
recovered for $\delta=0$. On the other hand, when $D=4$ and the NUT
parameter $m_2$ is set to zero, one recovers the Kerr--Sen solution of
the previous section if the fields  are rescaled as $\phi\to -\Phi$ and $\tens{A}\to \tens{A}/2$, and the following transformation of coordinates, and parameters is performed:
\ba
x_1&\to& ir\,,\quad  x_2\to a\cos\theta\,,\quad  \psi_0\to t-a\varphi\,,\quad \psi_1\to \varphi/a\,,\nonumber\\
c_0&\to& a^2\,, \quad  2m_1s^2\to 2ib\,,\quad 
im_1\to b-M\,.
\ea

Let us also introduce the orthonormal basis 
\be\label{1-form basis}
\tens{e}^\mu = \frac{\tens{d}x_\mu}{\sqrt{Q_\mu}}\,,\quad 
\tens{e}^{\hat{\mu}}=\sqrt{Q_\mu}\Big(\tens{{\cal A}}_\mu-\sum_{\nu=1}^n\frac{2N_\nu s^2}{HU_\nu}\,\tens{{\cal A}}_\nu\Big)\,,\quad 
\tens{e}^0 = \sqrt{S}\Big(\tens{{\cal A}}-\sum_{\nu=1}^n\frac{2N_\nu s^2}{HU_\nu}\,\tens{{\cal A}}_\nu\Big)\,, 
\ee
in which the metric and the field strengths are written as
\ba
\tens{g}&=&\sum_{\mu=1}^n(\tens{e}^\mu \tens{e}^\mu+\tens{e}^{\hat{\mu}}\tens{e}^{\hat{\mu}})+\varepsilon \tens{e}^0\tens{e}^0\,,\nonumber\\
\tens{F}&=&\frac{c}{s}\sum_{\rho=1}^n H_\rho ~\tens{e}^\rho\wedge \tens{e}^{\hat{\rho}}\,, \nonumber\\
\tens{H}&=&-\Big(\sum_{\mu=1}^n\sqrt{Q_\mu}~\tens{e}^{\hat{\mu}}+\varepsilon\,\sqrt{S}~\tens{e}^0\Big)\wedge
        \Big(\sum_{\rho=1}^n H_\rho ~\tens{e}^\rho\wedge \tens{e}^{\hat{\rho}}\Big)\,,
\ea
where we have denoted $H_\mu = \partial_\mu \ln H$. The inverse frame is given by 
\begin{align}\label{vector basis}
\tens{X}_\mu 
=& \sqrt{Q_\mu}\pa_{x_\mu} ~, \nonumber\\
\tens{X}_{\hat{\mu}} 
=& \sum_{k=0}^{n-1}\frac{(-1)^k x_\mu^{2(n-k-1)}}{U_\mu\sqrt{Q_\mu}}\,\pa_{\psi_k}
    +\frac{2N_\mu s^2}{U_\mu\sqrt{Q_\mu}}\,\pa_{\psi_0}
    +\frac{\varepsilon(-1)^nx_\mu^{-2}}{U_\mu\sqrt{Q_\mu}}\,\pa_{\psi_n} ~, \nonumber\\
\tens{X}_0
=& \frac{1}{\sqrt{S}A^{(n)}}\pa_{\psi_n} ~, 
\end{align} 
and the spin connection is calculated to be \cite{Chow:2008}\footnote{Note that there is a typo in the second line in (3.18) of \cite{Chow:2008}.}
\begin{align}\label{connection}
\tens{\omega}^\mu{}_\nu
=& -\frac{x_\nu\sqrt{Q_\nu}}{x_\mu^2-x_\nu^2}~\tens{e}^\mu
    -\frac{x_\mu\sqrt{Q_\mu}}{x_\mu^2-x_\nu^2}~\tens{e}^\nu ~~(\text{for $\mu\neq\nu$}) ~, \nonumber\\[0.3cm]
\tens{\omega}^\mu{}_{\hat{\mu}}
=& -H\partial_\mu\Big(\frac{\sqrt{Q_\mu}}{H}\Big)~\tens{e}^{\hat{\mu}}
    +\sum_{\rho\neq\mu}\frac{\sqrt{Q_\rho}}{2}\partial_\mu \ln (HU_\rho)~\tens{e}^{\hat{\rho}} 
+\varepsilon \sqrt{S}\Big(\frac{1}{x_\mu}+\frac{1}{2}\partial_\mu \ln H\Big)~\tens{e}^0
\nonumber\\[0.3cm]
\tens{\omega}^\mu{}_{\hat{\nu}}
=& \frac{\sqrt{Q_\nu}}{2}\partial_\mu \ln(HU_\nu)~\tens{e}^{\hat{\mu}}
    -\frac{x_\mu\sqrt{Q_\mu}}{x_\mu^2-x_\nu^2}~\tens{e}^{\hat{\nu}} ~~(\text{for $\mu\neq\nu$}) ~, \nonumber\\[0.3cm]
\tens{\omega}^{\hat{\mu}}{}_{\hat{\nu}}
=& -\frac{\sqrt{Q_\nu}}{2}\partial_\mu \ln(HU_\nu)~\tens{e}^\mu
    +\frac{\sqrt{Q_\mu}}{2}\partial_\nu \ln(HU_\mu)~\tens{e}^\nu ~~(\text{for $\mu\neq\nu$}) ~, \nonumber\\[0.3cm]
\tens{\omega}^\mu{}_0
=& \sqrt{S}\Big(\frac{1}{x_\mu}+\frac{1}{2}\partial_\mu \ln H\Big)~\tens{e}^{\hat{\mu}}
    -\frac{\sqrt{Q_\mu}}{x_\mu}~\tens{e}^0 ~, \nonumber\\[0.3cm]
\tens{\omega}^{\hat{\mu}}{}_0
=& -\sqrt{S}\Big(\frac{1}{x_\mu}+\frac{1}{2}\partial_\mu \ln H\Big)~\tens{e}^\mu ~. 
\end{align}

\subsection{Hidden symmetries}
The metric \eq{Chow solution} possesses $n+\eps$ obvious isometries $\pa_{\psi_k}$ ($k=0,\dots, n-1+\eps$).
Our claim is that in addition to these Killing vectors, the metric possesses a GCCKY 2-form $\tens{h}$,
\begin{align}
\tens{h} = \sum_{\mu=1}^n x_\mu ~\tens{e}^\mu\wedge
\tens{e}^{\hat{\mu}}\,, \label{canh}
\end{align}
which represents a natural generalization of the PCKY tensor of the Kerr--NUT spacetime \cite{KubiznakFrolov:2007} with respect to the following torsion:
\ba
\tens{T}= -\sum_{\mu=1}^n\sum_{\substack{\nu=1\\ \nu\neq\mu}}^n \sqrt{Q_\mu}H_\nu 
       ~\tens{e}^{\hat{\mu}\nu\hat{\nu}} 
      -\varepsilon\sum_{\mu=1}^n \sqrt{S}H_\mu ~\tens{e}^{0\mu\hat{\mu}}
  +\,\varepsilon\sum_{\mu=1}^n \frac{f}{x_\mu} ~\tens{e}^{0\mu\hat{\mu}} ~, \label{torsion}
\ea
where $f$ is an arbitrary function. Notice that this torsion is unique in an even number of spacetime dimensions; the non-uniqueness in odd dimensions, expressed by function $f$, follows from the fact that the 2-form $\tens{h}$ is necessary degenerate in odd dimensions.
Using the orthonormal basis \eq{1-form basis} and the connection \eq{connection} one can verify that 
\ba
\nabla_{X_\mu}^T\tens{h}
&=& \sum_{\nu=1}^n\sqrt{Q_\nu} ~\tens{e}^\mu\wedge \tens{e}^{\hat{\nu}} 
    +\varepsilon\sqrt{S}~\tens{e}^\mu\wedge \tens{e}^0
    +\frac{\varepsilon}{2}\,f~\tens{e}^\mu\wedge \tens{e}^0 ~, \nonumber\\[0.2cm]
\nabla_{X_{\hat{\mu}}}^T\tens{h}
&=& \sum_{\substack{\nu=1\\ \nu\neq\mu}}^n\sqrt{Q_\nu}~\tens{e}^{\hat{\mu}}\wedge \tens{e}^{\hat{\nu}}
    +\varepsilon\sqrt{S}~\tens{e}^{\hat{\mu}}\wedge \tens{e}^0
    +\frac{\varepsilon}{2}\,f~\tens{e}^{\hat{\mu}}\wedge \tens{e}^0 ~, \nonumber\\[0.2cm]
\nabla_{X_0}^T\tens{h}
&=& -\sum_{\rho=1}^n\sqrt{Q_\rho} ~\tens{e}^{\hat{\rho}}\wedge \tens{e}^0 ~.
\ea
and also
\be
\tens{\xi}=-\frac{1}{D-1}\,\delt\tens{h}
=\sum_{\mu=1}^n\sqrt{Q_\mu}~\tens{e}^{\hat{\mu}} +\eps\left(\sqrt{S}+\frac{f}{2}\right)\tens{e}^0\,.
\ee
It is then easy to prove that $\tens{h}$ obeys \eq{PCKY} for any vector field $\tens{X}$ and hence it is a GCCKY 2-form. 

Let us note that if we choose $f=0$, the torsion $\tens{T}$ becomes very natural and can be identified with the 3-form field strength $\tens{H}$. In that case we also have 
\be
\tens{T} = \tens{H}=-\frac{s}{c}\,\tens{F}\wedge \tens{\xi}\,.
\ee 

In any case, the GCCKY 2-form $\tens{h}$ gives rise to towers of
hidden symmetries as discussed in Section \ref{towers}. In particular, one obtains the tower of GCCKY $(2j)$-forms $\tens{h}^{(j)}$, $j=1,\dots,n-1$, and the following mutually Schouten commuting rank-2 irreducible Killing tensors \cite{Chow:2008}:
\begin{align}
\tens{K}^{(j)} = \sum_{\mu=1}^nA_\mu^{(j)}(\tens{e}^\mu \tens{e}^\mu+\tens{e}^{\hat{\mu}}\tens{e}^{\hat{\mu}})+\varepsilon A^{(j)}\tens{e}^0\tens{e}^0 ~,\qquad j=1,\dots,n-1\,.
\end{align}
Together with the Killing vector fields $\pa_{\psi_k}$ these Killing
tensors are responsible for complete integrability of geodesic motion
in the charged Kerr-NUT spacetimes, as discussed by Chow
\cite{Chow:2008}. Similarly, the  GCCKY tensors $\tens{h}^{(j)}$ are
responsible for separability of the Dirac equation. This requires the
vanishing of the anomalies, explicitly demonstrated in Appendix \ref{anom}. 
We shall now demonstrate that these hidden symmetries allow one to
separate the scalar and Dirac test fields in the charged Kerr-NUT
background. For simplicity, we consider only uncharged fields; the calculations extend results demonstrated in 
\cite{FrolovEtal:2007, OotaYasui:2008}.

\subsection{Separability of the scalar equation}
As in the four-dimensional case, let us consider a new scalar field $\varphi$ which in string frame obeys the following `massless' equation:\footnote{%
This equation is equivalent to the massless Klein--Gordon equation in the Einstein frame
$\square_E\varphi = 0$\,,
which was proved to separate by Chow \cite{Chow:2008}.
}
\begin{align}\label{scalar}
\square\varphi +\sqrt{\frac{D-2}{2}}\nabla_a\varphi\nabla^a\phi=0 ~,
\end{align}
where the background scalar field $\phi$ is given by \eq{Chow solution},
$\phi=\sqrt{\frac{2}{D-2}}\ln H$. This equation can be written as
\begin{equation}
\square\varphi +\sum_{\mu=1}^{n} H_\mu Q_\mu \frac{\partial \varphi}{\partial x_\mu}=0\,,
\end{equation}
and, using the basis \eq{vector basis}, it takes the following explicit form:
\begin{eqnarray}\label{scalar explicit}
\sum_{\mu=1}^{n} \frac{1}{U_\mu} \Biggl\{ X_\mu \frac{\partial^2 \varphi}{\partial x_\mu^2}\!&+&\!X_\mu^{'} \frac{\partial \varphi}{\partial x_\mu}
\!+\!\frac{1}{X_\mu} \left[\, \sum_{k=0}^{n-1} (-1)^k x_{\mu}^{2(n\!-\!k\!-\!1)} \frac{\partial}{\partial \psi_k}+2 N_\mu s^2\! \frac{\partial}{\partial \psi_0} 
+\varepsilon  \frac{(-1)^n}{x_{\mu}^2} \frac{\partial}{\partial \psi_n} \right]^2 \!\!\varphi
\nonumber\\ 
&+& \varepsilon \left[ \frac{(-1)^{n-1}}{\tilde{c} x_\mu^2} \frac{\partial^2 \varphi}{\partial \psi_n^2}+\frac{X_\mu}{x_\mu} \frac{\partial \varphi}{\partial x_\mu} \right] \Biggr\}=0\,.
\end{eqnarray}
This equation allows a multiplicative separation of variables
\begin{equation}\label{ansatz}
\varphi=\prod_{\mu=1}^{n} R_\mu(x_\mu) \prod_{k=0}^{n-1+\varepsilon} e^{i p_k \psi_k}\,.
\end{equation}
Indeed, plugging this ansatz into Eq. \eq{scalar explicit}, it assumes the form
\begin{equation}\label{separate}
\sum_{\mu=1}^{n} \frac{G_\mu}{U_\mu} \varphi=0\,,
\end{equation}
where $G_\mu$ is a function of $x_\mu$ only
\begin{equation}
G_\mu = X_\mu \frac{{R_\mu^{''}}}{R_\mu}
+\left( X_\mu^{'}+\varepsilon \frac{X_\mu}{x_\mu} \right)
\frac{R_\mu^{'}}{R_\mu}
-\frac{W_\mu^2}{X_\mu}+\varepsilon \frac{(-1)^n p_n^2}{\tilde{c} x_\mu^2}\,,
\end{equation}
and
\begin{equation}
W_\mu = \sum_{k=0}^{n-1} (-1)^k x_{\mu}^{2(n-k-1)} p_k+2 N_\mu s^2 p_0+\varepsilon \frac{(-1)^n p_n}{x_\mu^2}\,.
\end{equation}
The general solution of \eq{separate} is 
\begin{equation}
G_{\mu}=\sum_{j=1}^{n-1} k_j x_\mu^{2(n-1-j)}\,,
\end{equation}
where $k_j$ are arbitrary constants. Hence, the functions $R_\mu$ satisfy the ordinary second
order differential equations
\begin{equation}
(X_\mu R_\mu^{'})^{'}+\varepsilon \frac{X_\mu}{x_\mu} R_\mu^{'}-
\left(\frac{W_\mu^2}{X_\mu}+\sum_{j=1}^{n-1} k_j x_\mu^{2(n-1-j)}
-\varepsilon \frac{(-1)^n p_n^2}{\tilde{c} x_\mu^2} \right)R_\mu=0\,,
\end{equation}
and we have shown that the scalar field equation \eq{scalar} admits the multiplicative separation of variables \eq{ansatz}.

\subsection{Separability of the Dirac equation}
Finally, we demonstrate separability of the torsion modified Dirac
equation. For the time being, we will work with the torsion
\eq{torsion}, including an arbitrary function in odd spacetime
dimensions.\footnote{ We will find in Appendix A that we must specialize to the case
$\tens{T}=\tens{H}$, that is,  $f=0$, in order both anomalies \eq{Acl} and \eq{Aq} vanish.
However, separability of the Dirac equation can be demonstrated for other choices of $f$ as well.
} For simplicity we consider an uncharged field for which the Dirac equation reads
\be\label{DiracCH}
\bigl(\gamma^aD_a+m\bigr)\Psi=0\,,\quad\  D_a=X_a+\frac{1}{4}\gamma^b\gamma^c\omega_{bc}(X_a)-\frac{1}{24}\gamma^b\gamma^cT_{abc}\,.
\ee
Using the connection \eq{connection}, the inverse basis \eq{vector basis} and the torsion \eq{torsion} this equation takes the following explicit form:  
\begin{align}
& \Bigg\{\,
   \sum_{\mu=1}\gamma^\mu\sqrt{Q_\mu}
    \Bigg[\frac{\partial}{\partial x_\mu}+\frac{X_\mu^\prime}{4X_\mu}
          +\frac{\varepsilon}{2x_\mu}+\frac{1}{2}\sum_{\nu\neq\mu}\frac{x_\mu}{x_\mu^2-x_\nu^2}\Bigg] \nonumber\\
& ~~~+\sum_{\mu=1}^n\gamma^{\hat{\mu}}\sqrt{Q_\mu}
     \Bigg[\sum_{k=0}^{n-1+\varepsilon}\frac{(-1)^kx_\mu^{2(n-k-1)}}{X_\mu}\frac{\partial}{\partial\psi_k}
          +\frac{2N_\mu s^2}{X_\mu}\frac{\partial}{\partial\psi_0}
    +\frac{1}{2}\sum_{\substack{\nu=1\\ \nu\neq\mu}}^n
    \frac{x_\nu}{x_\mu^2-x_\nu^2}(\gamma^\nu\gamma^{\hat{\nu}})\Bigg] \nonumber\\
& ~~~+\varepsilon\,\gamma^0\sqrt{S}\Bigg[
    \frac{1}{c}\,\frac{\partial}{\partial\psi_n}
    -\sum_{\mu=1}^n\frac{F}{x_\mu}(\gamma^\nu\gamma^{\hat{\nu}})\Bigg]+m\,\Bigg\}\,\widetilde{\Psi} ~=~ 0 ~,
\end{align}
where we have set $\Psi=\sqrt{H}\widetilde{\Psi}$, using the fact that 
\begin{align}
\Big(\frac{\partial}{\partial x_\mu}-\frac{H_\mu}{2}\Big)(\sqrt{H}\widetilde{\Psi})
= \sqrt{H}\frac{\partial\widetilde{\Psi}}{\partial x_\mu} ~,
\end{align}
and we define an arbitrary function $F$ by $F = 1/2+f(4\sqrt{S})^{-1}$.  

Let us use the following representation of $\gamma$-matrices: $\{ \gamma^a, \gamma^b \} = 2 \delta^{ab}$,
\begin{eqnarray} \label{EQ14}
\gamma^{\mu}&=& 
\underbrace{\sigma_3 \otimes \sigma_3 \otimes \cdots \otimes \sigma_3}_{\mu-1}
\otimes \sigma_1 \otimes I \otimes \cdots \otimes I, \nonumber\\
\gamma^{\hat{\mu}}&=& 
\underbrace{\sigma_3 \otimes \sigma_3 \otimes \cdots \otimes \sigma_3}_{\mu-1} 
\otimes \sigma_2 \otimes I \otimes \cdots \otimes I, \nonumber\\
\gamma^0&=&\sigma_3 \otimes \sigma_3 \otimes \cdots \otimes \sigma_3\,,
\end{eqnarray} 
where $I$ is the $2 \times 2$ identity matrix and $\sigma_i$ are the Pauli matrices.
In this representation, we write the $2^n$ components of the spinor field as
$\Psi_{\epsilon_1 \epsilon_2  \cdots  \,\epsilon_n }~(\epsilon_{\mu}=\pm 1)$, and
it follows that
\begin{eqnarray} \label{EQ15}
(\gamma^{\mu}\Psi)_{\epsilon_1 \epsilon_2  \cdots  \epsilon_n }&=& \left(
\prod_{\nu=1}^{\mu-1}\epsilon_{\nu}\right)
\Psi_{\epsilon_1 \cdots  \epsilon_{\mu-1} (-\epsilon_{\mu})
\epsilon_{\mu+1} \cdots \epsilon_n },\nonumber\\
(\gamma^{\hat{\mu}}\Psi)_{\epsilon_1 \epsilon_2  \cdots  \epsilon_n } 
&=& -i \epsilon_{\mu}
\left(\prod_{\nu=1}^{\mu-1} \epsilon_{\nu}\right)
\Psi_{\epsilon_1 \cdots  \epsilon_{\mu-1} (-\epsilon_{\mu})
\epsilon_{\mu+1} \cdots \epsilon_n },\nonumber\\
(\gamma^0\Psi)_{\epsilon_1 \epsilon_2  \cdots  \epsilon_n } 
&=& \Big(\prod_{\rho=1}^n\epsilon_\rho\Big)
\Psi_{\epsilon_1 \cdots \epsilon_n }~.
\end{eqnarray} 
We consider the separable solution
\begin{align}
\widetilde{\Psi} = \hat{\Psi}(x)\exp \Big(i\sum_{k=0}^{n-1+\varepsilon}p_k\psi_k\Big) ~,
\end{align}
where $p_k$ ($k=0,\dots,n-1+\varepsilon$) are arbitrary constants.
Using (\ref{EQ15}), we obtain
\begin{align}
& \Bigg\{\,
   \sum_{\mu=1}\sqrt{Q_\mu}\Big(\prod_{\rho=1}^{\mu-1}\epsilon_\rho\Big)
    \Bigg[\frac{\partial}{\partial x_\mu}+\frac{X_\mu^\prime}{4X_\mu}
          +\frac{\varepsilon}{2x_\mu}+\frac{\epsilon_\mu Y_\mu}{X_\mu}
          +\frac{1}{2}\sum_{\nu\neq\mu}\frac{1-\epsilon_\mu\epsilon_\nu}{x_\mu+x_\nu}\Bigg] \nonumber\\
& ~~~+\varepsilon\,i\sqrt{S}\Big(\prod_{\rho=1}^{\mu-1}\epsilon_\rho\Big)
    \Bigg[\frac{p_n}{c}
    -\sum_{\mu=1}^n\frac{\epsilon_\mu F}{x_\mu}\Bigg]\,\Bigg\}
    \,\hat{\Psi}_{\epsilon_1\dots\epsilon_{\mu-1}(-\epsilon_\mu)\epsilon_{\mu+1}\dots \epsilon_n}
    +m\,\hat{\Psi}_{\epsilon_1\dots \epsilon_n} ~=~ 0 ~, \label{de1}
\end{align}
where
\begin{align}
Y_\mu = \sum_{k=0}^{n-1+\varepsilon}(-1)^kx_\mu^{2(n-k-1)}p_k+2N_\mu s^2 p_0 ~.
\end{align}
Following further \cite{OotaYasui:2008} we set\footnote{%
Note that there are some mistakes in \cite{OotaYasui:2008}. In Eq. (24) one needs to add a term $1/(2x_\mu)$ and  
replace the coefficient standing by $X_\mu^\prime/X_\mu$ by $1/4$. In Eq. (39) one should have
\begin{align*}
q_{-1} = \frac{i}{\sqrt{c}}(-1)^nN_n\,,\quad 
q_{-2} = \frac{i}{2}(-1)^{n-1}\sqrt{c} ~.
\end{align*}
}
\begin{align}
\hat{\Psi}_{\epsilon_1\dots \epsilon_n}(x) = 
\Bigg(\prod_{1\leq\mu<\nu\leq n}\frac{1}{\sqrt{x_\mu+\epsilon_\mu\epsilon_\nu x_\nu}}\Bigg)
\Bigg(\prod_{\mu=1}^n\chi^{(\mu)}_{\epsilon_\mu}(x_\mu)\Bigg) ~.
\end{align}
Thereafter,  we have the following equation
following from (\ref{de1}):
\begin{align}
\sum_{\mu=1}^n 
\frac{P^{(\mu)}_{\epsilon_\mu}(x_\mu)}{\prod_{\substack{\nu=1\\ \nu\neq\mu}}^n(\epsilon_\mu x_\mu-\epsilon_\nu x_\nu)}
+\frac{\varepsilon\,i\sqrt{c}}{\prod_{\rho=1}^n(\epsilon_\rho x_\rho)}
\Big(-\sum_{\mu=1}^n\frac{F}{\epsilon_\mu x_\mu}+\frac{p_n}{c}\Big)+m=0 ~, \label{EQ20}
\end{align}
where 
\begin{align}
P^{(\mu)}_{\epsilon_\mu} = 
(-1)^\mu(\epsilon_\mu)^{n-\mu}\sqrt{(-1)^{\mu-1}X_\mu}\frac{1}{\chi^{(\mu)}_{\epsilon_\mu}}
\Big(\frac{d}{dx_\mu}+\frac{X_\mu^\prime}{4X_\mu}
+\frac{\epsilon_\mu Y_\mu}{X_\mu}\Big)\chi^{(\mu)}_{-\epsilon_\mu} ~, 
\end{align}
are functions of $x_\mu$ only.

In order to satisfy (\ref{EQ20}) $P^{(\mu)}_{\epsilon_\mu}$ must assume the form
\begin{align}
P^{(\mu)}_{\epsilon_\mu}(x_\mu) = Q(\epsilon_\mu x_\mu)\,,
\end{align}
where (a) in an even dimension ($\varepsilon=0$) one has
\begin{align}
Q(y)=-m y^{n-1}+\sum_{j=0}^{n-2} q_j y^j\,,
\end{align}
whereas (b) in an odd dimension ($\varepsilon=1$) $Q$ is a solution of 
\begin{align}
\sum_{\mu=1}^n 
\frac{Q(y_\mu)}{\prod_{\substack{\nu=1\\ \nu\neq\mu}}^n(y_\mu-y_\nu)}
+\frac{i\sqrt{c}}{\prod_{\rho=1}^n y_\rho}
\Big(-\sum_{\mu=1}^n\frac{F}{y_\mu}+\frac{p_n}{c}\Big)+m=0 ~.
\end{align}
In particular, for $F=1/2$ (which corresponds to the natural torsion $\tens{T}=\tens{H}$) we have 
\begin{align}
Q(y)=\sum_{j=-2}^{n-1} q_j y^j ~,\quad \ q_{n-1}=-m\,,\ \ q_{-1}=\frac{i}{\sqrt{c}}(-1)^n p_n\,,\ \ q_{-2}=\frac{i}{2} (-1)^{n-1} \sqrt{c}\,.
\end{align}
In both cases parameters $q_j$ ($j=0,\dots,n-2$) are arbitrary.

Let us summarize our result. We have proved that the torsion modified
Dirac equation \eq{DiracCH} in the charged Kerr-NUT spacetime \eq{Chow
  solution} allows separation of variables 
\begin{align}
\Psi_{\epsilon_1\dots \epsilon_n} = 
\sqrt{H}
\Bigg(\prod_{1\leq\mu<\nu\leq n}\frac{1}{\sqrt{x_\mu+\epsilon_\mu\epsilon_\nu x_\nu}}\Bigg)
\Bigg(\prod_{\mu=1}^n\chi^{(\mu)}_{\epsilon_\mu}(x_\mu)\Bigg)
\exp \Bigg(i\sum_{k=0}^{n-1+\varepsilon}p_k\psi_k\Bigg) ~,
\end{align}
where functions $\chi^{(\mu)}_{\epsilon_{\mu}}$ 
satisfy the (coupled) ordinary first order differential equations
\begin{align}
\left(\frac{d}{dx_{\mu}}+\frac{1}{4} \frac{X_{\mu}^{'}}{X_{\mu}}
+\frac{\epsilon_{\mu} Y_{\mu}}{X_{\mu}}
\right) \chi^{(\mu)}_{-\epsilon_{\mu}}-
\frac{(-1)^{\mu-1}(\epsilon_{\mu})^{n-\mu} Q(\epsilon_{\mu} x_{\mu})}
{\sqrt{(-1)^{\mu-1} X_{\mu}}}  \chi^{(\mu)}_{\epsilon_{\mu}}=0 ~.
\end{align}

The demonstrated separation is justified by the existence of the GCCKY
2-form $\tens{h}$. As in the four-dimensional case it is demonstrated in Appendix \ref{anom} 
that for all GCCKY $(2j)$-forms $\tens{h}^{(j)}$ both anomalies
\eq{Acl} and \eq{Aq} vanish and the corresponding operators of Lemma \ref{GCCKYoplem} provide symmetry operators which commute with the modified Dirac operator.

\section{Conclusions}
In this paper we have studied an extension of Killing--Yano symmetry in the presence of skew-symmetric torsion.
We have demonstrated, that when the torsion is an arbitrary 3-form, one obtains various torsion anomalies and the implications of the existence of
the generalized Killing--Yano symmetry are relatively weak. For example, contrary to the vacuum case neither complete integrability of geodesic motion nor separability of test field equations are implied in general.
However, in the spacetimes where there is a natural 3-form, obeying the appropriate field equations,
these anomalies may disappear and the concept of generalized
Killing--Yano symmetry may become very powerful. This is for example
the case of the black hole of minimal supergravity where the torsion
is identified with the dual of Maxwell field \cite{KubiznakEtal:2009b},
or, as demonstrated in this paper, the case of the Kerr--Sen solution
of effective string theory and its higher-dimensional generalizations
where the torsion is identified with the 3-form $\tens{H}$. In both
cases a choice of the torsion is very natural and the generalized
Killing--Yano symmetry carries non-trivial information about the
spacetime; for example it underlines complete integrability of
geodesic motion as well as separability of the scalar and Dirac equations. 

It is an interesting question whether the Killing--Yano symmetry and
its generalizations can provide new insights into the theory of black
holes beyond its many contributions to the vacuum theory.

\section*{Acknowledgments}
We wish to thank G.W.~Gibbons for reading the manuscript. 
D.K. is grateful to the Herschel Smith Postdoctoral Research Fellowship
at the University of Cambridge. The work of Y.Y. is supported by the Grant-in Aid for Scientific Research
No.21244003 from Japan Ministry of Education. He is also grateful for the hospitality of
DAMTP, University of Cambridge during his stay.

\appendix

\section{Vanishing of anomalies \label{anom}}
In this appendix we wish to justify (at least partially) the
separability of the modified Dirac equation  for $\tens{T}=\tens{H}$ in charged Kerr-NUT spacetimes by proving that for all GCCKY $(2j)$-forms $\tens{h}^{(j)}$ both anomalies \eq{Acl} and \eq{Aq} vanish and hence the corresponding operators  $M_{h^{(j)}}$, \eq{Mh}, give symmetry operators for the Dirac operator \eq{DiracT}.\footnote{%
In order to justify separability completely, one should additionally prove that all such operators mutually commute. Such a task 
is rather more difficult.
}
Recall that from \eq{canh} the GCCKY 2-form $\tens{h}^{(1)}=\tens{h}$ is given by
\begin{equation}
\tens{h}=\sum_{\mu=1}^{n} x_\mu e^\mu \wedge e^{\hat{\mu}}.
\end{equation}
The torsion 3-form is given by
\begin{equation}
\tens{T}=-\sum_{\mu \ne \nu} \sqrt{Q_\mu} H_\nu e^{\hat{\mu}} \wedge
e^\nu \wedge e^{\hat{\nu}}-\varepsilon \sum_{\mu} \sqrt{S} H_\mu e^0
\wedge e^\mu \wedge e^{\hat{\mu}},
\end{equation}
where we have taken the arbitrary function $f$ in \eq{torsion} to vanish. From \eq{Acl} and \eq{Aq}, we have
\begin{eqnarray}
\tens{A}_{(cl)} (\tens{h}^{(j)})&=&
-\frac{\tens{T} \wedge \tens{\delta}^T \tens{h}^{(j)}}{D-2j+1}
-\frac{1}{2} \tens{d}\tens{T} \underset{1}{\wedge} \tens{h}^{(j)}, \nonumber\\
\tens{A}_{(q)} (\tens{h}^{(j)})&=&
\frac{\tens{\delta} \tens{\delta}^T \tens{h}^{(j)}}{D-2j+1}
+\frac{1}{12} \tens{d}\tens{T} \underset{3}{\wedge} \tens{h}^{(j)}.
\end{eqnarray}
For $j=1$ it is easy to confirm by using the explicit form of
$\tens{h}$ that
\begin{eqnarray}
\tens{T} \wedge \tens{\delta}^T \tens{h}=\tens{d}\tens{T} \underset{1}{\wedge} \tens{h}=0, \\
\tens{\delta} \tens{\delta}^T \tens{h}=\tens{d}\tens{T} \underset{3}{\wedge} \tens{h}=0,
\end{eqnarray}
which leads to
\begin{equation}
\tens{A}_{(cl)} (\tens{h})=\tens{A}_{(q)} (\tens{h})=0.
\end{equation}
In general, provided the classical anomaly vanishes for the GCCKY forms $\tens{h}_1,~\tens{h}_2$,
then it also vanishes for $\tens{h}_1 \wedge \tens{h}_2$.
Hence we have $\tens{A}_{(cl)} (\tens{h}^{(j)})=0.$ 

In order to show $\tens{A}_{(q)} (\tens{h}^{(j)})=0$
we prove the following statement
\be
\tens{\delta} \tens{\delta}^T \tens{h}^{(j)} =0 . \label{deldelh}
\ee
Using the formula
\begin{equation}
\frac{1}{D-(p_1+p_2)+1} \tens{\delta}^T (\tens{h}_1 \wedge \tens{h}_2)
=\frac{1}{D-p_1+1} \tens{\delta}^T \tens{h}_1 \wedge \tens{h}_2
+\frac{(-1)^{p_1}}{D-p_2+1}\tens{h}_1 \wedge \tens{\delta}^T \tens{h}_2
\end{equation}
for GCCKY $p_i$-forms $\tens{h}_1$ and $\tens{h}_2$, we obtain
\begin{equation}
\tens{\delta}^T \tens{h}^{(j)}=\frac{(D-2 j+1)j}{D-1} \tens{\delta}^T \tens{h} \wedge \tens{h}^{(j-1)}.
\end{equation}
Further we calculate
\begin{equation}
\tens{\delta}(\tens{\delta}^T \tens{h} \wedge \tens{h}^{(j-1)})=\tens{\delta}(\tens{\delta}^T \tens{h} \wedge \tens{h}^{(j-2)}) \wedge \tens{h}
+\tens{h}^{(j-2)} \wedge I_1 + 2(j-2)\tens{h}^{(j-3)} \wedge \tens{\delta}^T \tens{h} \wedge I_2,
\end{equation} 
where
\begin{eqnarray}
I_1 &=& \nabla_{X_a} \tens{\delta}^T \tens{h} \wedge X_a \hook \tens{h}- X_a \hook \tens{\delta}^T \tens{h} \wedge \nabla_{X_a} \tens{h}
-\tens{\delta}^T \tens{h} \wedge \tens{\delta} \tens{h}, \nonumber\\
I_2 &=& X_a \hook \tens{h} \wedge \nabla_{X_a} \tens{h}.
\end{eqnarray}
By a direct calculation we have $I_1=0$ and $I_2$ is proportional to the factor $\tens{\delta}^T \tens{h}$,
\begin{equation}
I_2= \frac{1}{D-1} \tens{\delta}^T \tens{h} \wedge \left( \sum_{\mu=1}^n(2 +x_\mu H_\mu)x_\mu e^\mu \wedge e^{\hat{\mu}} \right),
\end{equation}
so that 
\begin{equation}
\tens{\delta}(\tens{\delta}^T \tens{h} \wedge \tens{h}^{(j-1)})=\tens{\delta}(\tens{\delta}^T \tens{h} \wedge \tens{h}^{(j-2)}) \wedge \tens{h}.
\end{equation}
By induction, we therefore establish \eq{deldelh}. Finally, one may
show that
\begin{equation}
\tens{d}\tens{T} \underset{3}{\wedge} \tens{h}^{(j)}=0
\end{equation}
by a direct calculation using
$\tens{d}\tens{T}=\tens{d}\tens{H}=-\tens{F}\wedge\tens{F}$. Making
use of \eq{deldelh} we conclude $\tens{A}_{(q)} (\tens{h}^{(j)})=0$.

Next we show that the tower of GKY forms, $\tens{f}^{(j)}=*\tens{h}^{(j)}$, also satisfies 
the anomaly free condition,
\begin{equation}
\tens{A}_{(cl)} (\tens{f}^{(j)})=\tens{A}_{(q)} (\tens{f}^{(j)})=0,
\end{equation}
where
\begin{eqnarray}
\tens{A}_{(cl)} (\tens{f}^{(j)})&=&\frac{\tens{d} \tens{d}^T \tens{f}^{(j)}}{p+1}-\frac{1}{2} \tens{d}\tens{T} \underset{1}{\wedge} \tens{f}^{(j)}, \nonumber\\
\tens{A}_{(q)} (\tens{f}^{(j)})&=&-\frac{1}{6(p+1)} \tens{T} \underset{3}{\wedge} \tens{d}^T\tens{f}^{(j)}+\frac{1}{12} \tens{d}\tens{T} \underset{3}{\wedge} \tens{f}^{(j)}.
\end{eqnarray}
In general, the following result holds for contracted wedge products
\begin{align}
\tens{\alpha} \underset{r}{\wedge} \tens{*\beta} = (-1)^{p(q+r+1)}\frac{r!}{(p-r)!} \tens{*} (\tens{\alpha}\underset{p-r}{\wedge} \tens{\beta}) ~,
\end{align}
where $\tens{\alpha}$ and $\tens{\beta}$ are $p$ and $q$ forms, respectively. In particular, we have
\begin{eqnarray}
\tens{d}\tens{T} \underset{3}{\wedge} \tens{f}^{(j)} &=& 
3! * ( \tens{d}\tens{T} \underset{1}{\wedge} \tens{h}^{(j)})=0,~\nonumber\\
\tens{d}\tens{T} \underset{1}{\wedge} \tens{f}^{(j)} &=& 
\frac{1}{3!} * ( \tens{d}\tens{T} \underset{3}{\wedge} \tens{h}^{(j)})=0.
\end{eqnarray}
Further we find
\begin{eqnarray}
\tens{d} \tens{d}^T \tens{f}^{(j)} &=&-* \tens{\delta} \tens{\delta}^T \tens{h}^{(j)}=0\,,~\nonumber\\
\tens{T} \underset{3}{\wedge} \tens{d}^T \tens{f}^{(j)}&=& 3! *( \tens{\delta}^T \tens{h}^{(j)} \wedge \tens{T})=0\,.
\end{eqnarray}
Thus, we conclude the anomaly vanishes for GKY forms.

%\bibliography{Databaze}
%\bibliographystyle{JHEP}

\providecommand{\href}[2]{#2}\begingroup\raggedright\endgroup

\end{document}